\documentclass[12pt, draftclsnofoot, onecolumn]{IEEEtran}

\usepackage{cite}
\usepackage{algorithmic}
\usepackage{array}
\usepackage{bbm}
\usepackage{amsfonts}
\usepackage{graphicx}
\usepackage[hidelinks]{hyperref}
\usepackage{xcolor}
\hypersetup{
	colorlinks  = true,
	linkcolor	={red!100!black},
	citecolor	={blue!100!black},
	urlcolor	={blue!100!black}
}
\usepackage{mathtools}
\usepackage{subfigure,epsfig}
\usepackage{amsmath, amsthm, amssymb}
\usepackage[outdir=./]{epstopdf}
\usepackage{dsfont}
\usepackage{mathrsfs}
\newtheorem{Theorem}{Theorem}
\newtheorem{Corollary}{Corollary}

\newtheorem{Proposition}{Proposition}

\def\figref#1{Fig.\,\ref{#1}}%
\newlength{\figwidth}
\begin{document}
\title{Stochastic Geometry Analysis of Sojourn Time in Multi-Tier Cellular Networks}
\author{Mohammad Salehi and Ekram Hossain\thanks{The authors are with the Department of Electrical and Computer Engineering at the University of Manitoba, Canada 
		(emails: salehim@myumanitoba.ca, Ekram.Hossain@umanitoba.ca). \textbf{E. Hossain} is the corresponding author.
}}

\maketitle

\begin{abstract}	
Impact of mobility will be increasingly important in future generation wireless services and the related challenges will need to be addressed. Sojourn time, the time duration that a mobile user stays within a cell, is a mobility-aware parameter that can significantly impact the  performance of mobile users and it can also be exploited to improve resource allocation and mobility management methods in the network. In this paper, we derive the distribution and mean of the sojourn time in multi-tier cellular networks, where spatial distribution of base stations (BSs) in each tier follows an independent homogeneous Poisson point process (PPP). To obtain the sojourn time distribution in multi-tier cellular networks with maximum biased averaged received power association, we derive the linear contact distribution function and chord length distribution of each tier. We also study the relation between mean sojourn time and other mobility-related performance metrics. We show that the mean sojourn time is inversely proportional to the handoff rate, and the complementary cumulative distribution function (CCDF) of sojourn time is bounded from above by the complement of the handoff probability. Moreover, we study the impact of user velocity and network parameters on the sojourn time.
\end{abstract}


\newpage
\section{Introduction}

\subsection{Background and Related Work}

The next generations of cellular wireless  networks are expected to support communications for highly mobile users and devices \cite{tabassum2019} with applications in new vertical sectors such as railway, unmanned aerial vehicle (UAV), and autonomous car. Therefore, addressing the mobility related challenges is necessary for the development of the next generation cellular networks. Impact of user/device mobility on its performance in cellular networks can be measured through mobility-aware performance metrics such as handoff rate, handoff probability, and sojourn time \cite{tabassum2019}. Sojourn time (or dwell time), time duration that a mobile user stays within a cell, is a key network parameter which allows studying other important network parameters such as channel occupancy time, new call and handoff call dropping probabilities \cite{corral2011}. Therefore, it is imperative to incorporate the sojourn time distribution in resource allocation and mobility management for improving the network performance.   In general, modeling and analysis of mobility-related parameters and performances is however challenging in multi-tier (or heterogeneous) cellular networks (e.g. a two-tier macrocell-small cell network) since it needs to consider different aspects such as how to model the distributions of base stations (BSs) at the different tiers, how to model the user mobility and traffic at the different tiers, and how to model the radio access network performance at the different tiers~\cite{rrm-book-ekram}. 

In this above context, \cite{lin2013} derived the sojourn time distribution for the hexagonal (deterministic) cellular networks and Poisson (random) cellular networks, where the BSs are distributed according to a homogeneous Poisson point process (PPP). In \cite{shin2016}, mean sojourn time of two-tier cellular networks was approximately derived, where the coverage areas of macro cells and small cells have regular shapes (circles) and within each macro cell multiple small cells are irregularly deployed. \cite{Xu2017,Hong2015} derived the mean sojourn time in small cells of two-tier cellular networks. The BSs of each tier are distributed following an independent homogeneous PPP. \cite{Lee2014} also derived the mean sojourn time in two-tier cellular networks. However, it was assumed that a handoff occurs only when the mobile user crosses the boundary of a macro cell. Therefore, the mean sojourn time in \cite{Lee2014} is similar to that in a single-tier network as in \cite{lin2013}. 

Moreover, the handoff rate, i.e. the expected number of handoffs in unit time, was derived in \cite{lin2013} for single-tier Poisson cellular networks and in \cite{bao2015stochastic} for multi-tier Poisson cellular networks. The handoff probability, i.e. the probability that the mobile user handoffs to a new BS at the end of a movement period, was also studied in \cite{Sadr2015} and \cite{Hsueh2017} for single-tier and multi-tier Poisson networks, respectively. To derive the mean sojourn time (or distribution of the sojourn time), \cite{lin2013,Hong2015,Lee2014} used the {\em chord length distribution} (or {\em linear contact distribution function}) of Poisson Voronoi cells. However, in multi-tier networks with different transmission power and bias factor for each tier, we need the chord length distribution (or linear contact distribution function) of weighted Poisson Voronoi cells which is not available in the literature. For single-tier networks, \cite{tabassum2019}  used the handoff probability to derive the distribution of sojourn time in the cell where connection is initiated. In single-tier networks, since the Voronoi cells are convex, we can directly use the handoff probability to  derive the distribution of the sojourn time. However, in multi-tier networks, cells may not be convex. Therefore, the analytical method in \cite{tabassum2019} cannot be used for  multi-tier networks. 

A handoff is considered to be unnecessary when the dwell time of the mobile user in the new cell after the handoff is less than a predefined threshold. In \cite{Fu2013,Arshad2016}, handoff skipping schemes are employed to avoid unnecessary handoffs. Moreover, important system parameters such as channel occupancy time, new call and handoff call dropping probabilities depend on the sojourn time \cite{corral2011}. Therefore, sojourn time is fundamental for analysis and design of the mobile cellular networks. In \cite{Fu2013,corral2011,Fang,Khan,Thajchayapong}, different distributions such as exponential, Erlang, gamma, Pareto, and Weibull were used for modeling the sojourn time distribution. Due to the principal role of the sojourn time in mobility management and resource allocation, in this paper, {\em we derive the sojourn time distribution in multi-tier cellular networks}.    

\subsection{Contributions}
To analyze the sojourn time distribution in multi-tier scenarios with PPP distributed BSs, the existing works either assume that the mobile user is always associated to only one of the tiers, or only focus on the small tier (in two-tier scenarios). For both the cases, the results are no different from the single-tier scenarios. In single-tier networks with maximum  averaged received power association (nearest BS association), (Voronoi) cells are convex; however, in multi-tier networks with maximum biased averaged received power association, cells may not be convex depending on the transmission power and bias factor of each tier. Therefore, analysis of sojourn time of multi-tier cellular networks is more complicated compared to the single-tier networks. In this regard, the contributions of this paper can be summarized as follows:
\begin{itemize}
	\item We derive the distribution and mean of the sojourn time for multi-tier cellular networks. We show that the mean sojourn time is inversely proportional to velocity. We also study the impact of network parameters on the sojourn time.
	\item To obtain the analytical results, we derive the linear contact distribution and chord length distribution of each tier.
	\item We show that the mean sojourn time is inversely proportional to handoff rate. Also, using handoff rate and sojourn time, we calculate the ping-pong rate (i.e. rate of unnecessary handoffs) for each tier.
	\item We show that the complement of the handoff probability provides an upper bound for the complementary cumulative distribution function (CCDF) of the sojourn time. We also discuss the scenarios where the CCDF of the sojourn time is equal to the complement of the handoff probability.
\end{itemize}

The rest of this paper is organized as follows: In Section~II, the system model is presented. In Section~III, we state the methodology for deriving the analytical results. In Sections IV and V, we obtain the main results related to the distribution and mean of the sojourn time and also discuss the effects of network parameters. Numerical and simulation results are provided in Section~VI. Finally, in Section~VII, we conclude the paper. 

\section{System Model and Notations}
Consider a $K$-tier heterogeneous cellular network with $K$ classes of BSs and let $\mathcal{K}=\{1,2,...,K\}$. The spatial distribution of BSs of $k$-th tier, $k\in\{1,2,...,K\}$, follows an independent homogeneous PPP $\Phi_{k}$ of intensity $\lambda_{k}$. Different tiers of BSs transmit at different power levels. $P_k$ denotes the transmission power of the $k$-th tier BSs. 

Consider a typical mobile user which moves in a straight line with a constant velocity $v$. Due to the stationarity of the homogeneous PPP, i.e. its distribution is invariant under translation \cite{haenggi2012stochastic}, we can assume that the typical mobile user is located at the origin $o$ at time 0. Since homogeneous PPP is isotropic, i.e. its distribution is invariant under rotation with respect to the origin \cite{haenggi2012stochastic}, we can also assume that the typical mobile user moves along the positive $x$-axis. Therefore, at time $t$, the typical mobile user is located at $\text{x}(t)=\left(vt,0\right)$.

The mobile user is always associated to the BS which provides the maximum biased averaged received power. Let us denote the serving BS at time $t$ by $BS(t)$. Therefore, 
\begin{IEEEeqnarray}{rCl}
	BS(t) = \text{arg}\max\limits_{x\in\Phi_k, \forall k\in\mathcal{K}} B_k
	P_k \|\text{x}(t)-x\|^{-\alpha}, 
	\label{eq:association-policy}
\end{IEEEeqnarray}
where $B_k$ is the cell range expansion bias factor for tier-$k$, and $\alpha$ is the path-loss exponent. Let us denote the distance between $BS(0)$ and the mobile user at $\text{x}(t)$ by $r_0(t)$, i.e. $r_0(t)=\|BS(0)-\text{x}(t)\|$. Given that at time $t$ the mobile user is associated to a tier-$k$ BS at distance $r(t)$, from \eqref{eq:association-policy}, we have $\Phi_j\left( \mathcal{B}\left( \text{x}(t),\frac{r(t)}{\beta_{kj}} \right) \right)=0$, $\forall j\in\mathcal{K}$, where $\beta_{kj}=\left(\frac{B_k P_k}{B_j P_j}\right)^{1/\alpha}$, $\mathcal{B}(\text{x},r)$ denotes a ball with radius $r$ centered at $\text{x}$, and $\Phi_j(A)$ is the number of tier-$j$ BSs in set $A\subset\mathbb{R}^2$. For simplicity, we define $r_0 = r_0(0)$. 

A summary of the major notations is provided in Table \ref{Table1}.
\begin{table*}[!ht]
	\centering
	\caption{Summary of Notations}
	\label{Table1}
	\begin{tabular}{|p{2.0cm}|p{11cm}|}
		\hline
		{\bf Notation} & { \bf Description} 
		\vspace{3mm}
		\\ 
		\hline
		$\Phi_{k}$, $\lambda_{k}$ & PPP of tier-$k$ BSs, intensity of $\Phi_{k}$ 
		\\
		\hline
	    $P_k$ & Transmit power of the $k$-th tier BSs 
		\\
		\hline
		$\text{x}(t)$ & Location of the mobile user at time $t$
		\\ 
		\hline
		$v$ & Velocity of the mobile user    
		\\
		\hline
		$BS(t)$ & Serving BS of the mobile user at time $t$      
		\\
		\hline
		$B_k$ & Cell range  expansion (bias) factor for tier $k$ 
		\\
		\hline
		$\alpha$ & Path-loss exponent 
		\\
		\hline
		$r_0(t)$, $r_0$ & Distance between the initially serving BS and the mobile user at time $t$, $r_0(0)$   
		\\
		\hline
		$\beta_{kj}$ & $\left( \frac{B_k P_k}{B_j P_j} \right)^{1/\alpha}$    
		\\
		\hline
		$\mathcal{B}(\text{x},r)$ & Ball with radius $r$ centred at $\text{x}$   
		\\
		\hline
		$\tilde{S}$ & Sojourn time in the cell where connection is initiated
		\\
		\hline
		$S$ & Sojourn time
		\\
		\hline
		$H_k$ & Handoff rate from (to) a tier-$k$ cell to (from) any other cell in the network
		\\
		\hline
		$H$ & Handoff rate 
		\\
		\hline
	\end{tabular}
\end{table*}

\section{ Methodology of Analysis of Sojourn Time in Multi-Tier Cellular Networks}
\begin{figure}
	\parbox[c]{.5\textwidth}{%
		\centerline{\subfigure[Single-tier.]{
			\includegraphics[width=.5\textwidth]{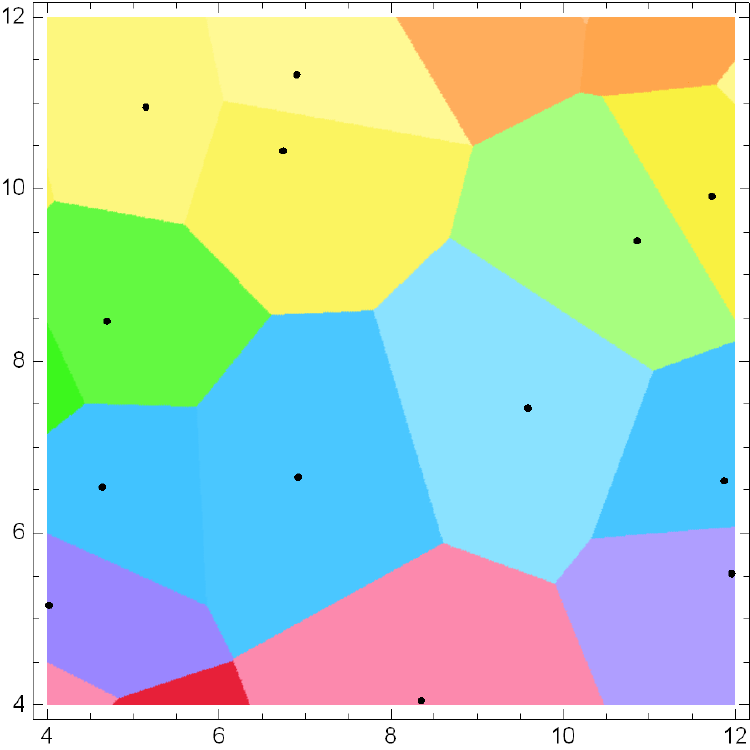}}}} 
	\parbox[c]{.5\textwidth}{%
		\centerline{\subfigure[Two-tier.]{
			\includegraphics[width=.5\textwidth]{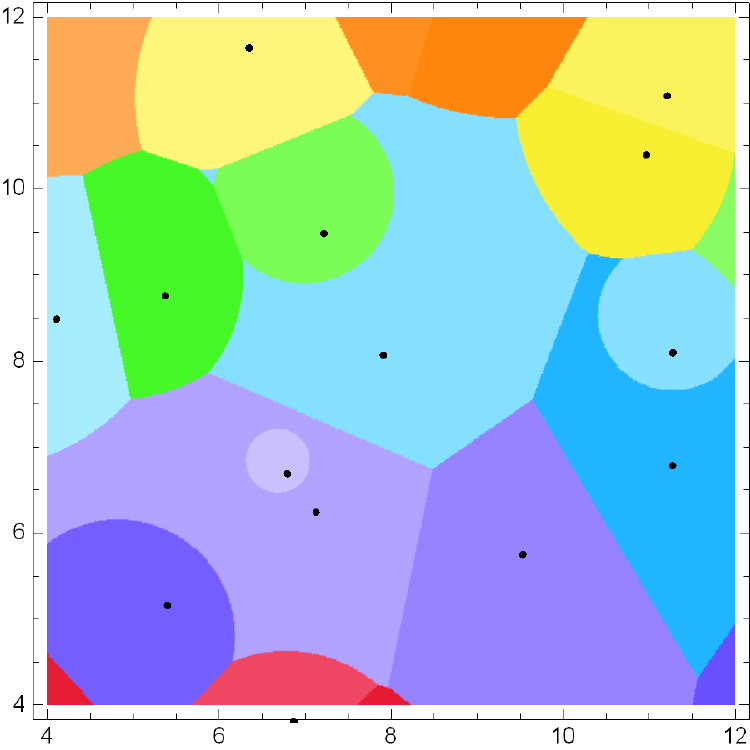}}}} 
	\caption{Voronoi cells in a single-tier and a two-tier cellular networks: (a) $\lambda=0.15$, (b) $\lambda_1=0.05$, 
	$\lambda_2=0.1$, and $\beta_{12}=2$. }
	\label{fig:convex_nonconvex}
\end{figure}
The sojourn time $S$ is the duration that the mobile user stays within a particular serving cell before it is handed over to another cell \cite{saad2016}. Analysis of sojourn time in multi-tier cellular networks consists of the following four  steps:
\begin{itemize}
	\item \textbf{Step 1}: Deriving the conditional distribution of the sojourn time in the cell where connection is initiated, given that the mobile user is initially associated to a tier-$k$ BS.
	\item \textbf{Step 2}: Deriving the linear contact distribution function, given that the mobile user is in a tier-$k$ cell at time 0. 
	\item \textbf{Step 3}: Obtaining the chord length distribution for tier $k$ using linear contact distribution function.
	\item \textbf{Step 4}: Deriving the distribution of the sojourn time $S$ for tier $k$. 
\end{itemize}

\subsection{Step 1 of Analysis}
First we focus on the distribution of the sojourn time in the cell where connection is initiated $\tilde{S}$. Specifically, we derive the CCDF of $\tilde{S}$, i.e.
\begin{IEEEeqnarray}{rCl}
	\bar{F}_{\tilde{S}}(T) = \mathbb{P}(\tilde{S}>T) 
	                       &=& \mathbb{P}\left(\text{no handoff occurs in the interval } [0,T]\right) \nonumber \\
	                       &=& \mathbb{P}\left( BS(t)=BS(0), \forall t\in(0,T]\right). 
	\label{eq-sojourn-time-def}
\end{IEEEeqnarray}
In single-tier cellular networks, Voronoi cells are convex \cite{moller2012lectures} (as shown in \figref{fig:convex_nonconvex}(a)). A set $C$ is convex if the line segment between any two points in $C$ lies in $C$ \cite{boyd2004convex}. Thus, in single-tier cellular networks, when the mobile user is connected to the same BS at time 0 and $T$, i.e. when $BS(T)=BS(0)$, the serving BS at any time between 0 and $T$ is also $BS(0)$, i.e. $BS(t)=BS(0)$, $\forall t\in(0,T)$. Hence, for single-tier cellular networks, \eqref{eq-sojourn-time-def} can be simplified as
\begin{IEEEeqnarray}{rCl}
	\bar{F}_{\tilde{S}}(T) = \mathbb{P}\left( BS(T)=BS(0) \right)\footnote{ $\mathbb{P}\left( BS(T) \neq BS(0) \right)$ is the probability that the mobile user is handed off  to a new BS after the movement period of $T$, and it is called handoff probability in the literature. According to \eqref{eq-sojourn-time-def-single-tier}, in single-tier networks, the CCDF of the sojourn time is equal to the complement of the handoff probability.}.  
	\label{eq-sojourn-time-def-single-tier}
\end{IEEEeqnarray} 
However, for multi-tier cellular networks, Voronoi cells may not be convex depending on the values of $\beta_{kj}$, $k,j\in\mathcal{K}$ (as shown in \figref{fig:convex_nonconvex}(b)). Therefore, even when $BS(T)=BS(0)$, there may exist a time $t$ between 0 and $T$ for which $BS(t) \neq BS(0)$. To derive the CCDF of $\tilde{S}$ for multi-tier cellular networks, we must use \eqref{eq-sojourn-time-def}, which makes the analysis of sojourn time in multi-tier cellular networks more complicated compared to the single-tier networks. Actually, single-tier scenario can be considered as a special case of multi-tier scenarios. Moreover, note that, \eqref{eq-sojourn-time-def-single-tier} provides an upper bound for \eqref{eq-sojourn-time-def}.

\begin{figure}
	\centering
	\includegraphics[width=.5\textwidth]{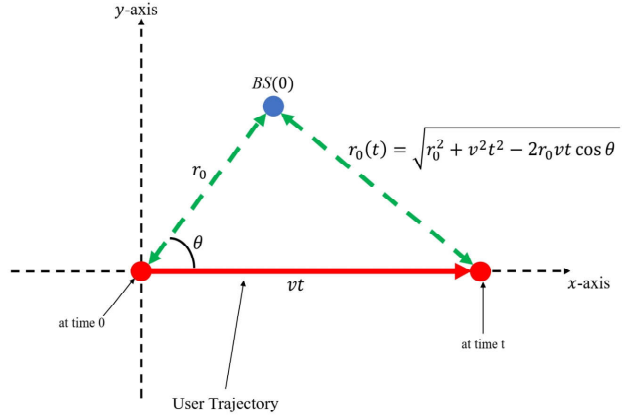}
	\caption{System model.}
	\label{fig:system_model}	
\end{figure}
Given that the mobile user is initially connected to a tier-$k$ BS, the CCDF of $\tilde{S}$ can be obtained by 
\begin{IEEEeqnarray}{rCl}
	\IEEEeqnarraymulticol{3}{l}{ 
		\bar{F}_{\tilde{S}}(T \mid \text{tier}=k) = \mathbb{P}\left( BS(t)=BS(0), \forall t\in(0,T] \mid \text{tier}=k \right)
	} \nonumber 
	\\
	 &=& \frac{1}{\pi} \int_0^{\infty} \int_0^{\pi} \mathbb{P}\left( BS(t)=BS(0), \forall t\in(0,T] \mid r_0, \theta, \text{tier}=k \right) f_R(r_0 \mid \text{tier}=k) {\rm d}\theta {\rm d}r_0,
	 \label{eq:Step1-1}
\end{IEEEeqnarray}
where $\theta$ is the angle between the serving BS at time 0 and direction of the movement (as shown in \figref{fig:system_model}). $\theta$ is uniformly distributed in $[0,\pi]$. $f_R(r_0 \mid \text{tier}=k)$ is the probability density function (PDF) of the serving link distance at time 0, given that $BS(0)$ belongs to tier-$k$. According to \cite{jo2012heterogeneous}, 
\begin{IEEEeqnarray}{rCl}
	f_R(r_0 \mid \text{tier}=k) = \frac{ 1 }{ \mathbb{P}(\text{tier}=k) } 2\lambda_k\pi r_0 \exp\left\{ -\sum_{j\in\mathcal{K}} \lambda_j \pi \beta_{jk}^2 r_0^2 \right\},
	\label{eq:PDF_r0}
\end{IEEEeqnarray}
where $\mathbb{P}(\text{tier}=k)$ is the probability that $BS(0)$ belongs to tier-$k$ which is given by \cite{jo2012heterogeneous}:
\begin{IEEEeqnarray}{rCl}
	\mathbb{P}(\text{tier}=k) = \frac{ \lambda_k }{ \sum_{j\in\mathcal{K}} \lambda_j \beta_{jk}^2 }.
	\label{eq:tier_k}
\end{IEEEeqnarray}

\begin{figure}
	\parbox[c]{.32\textwidth}{%
		\centerline{\subfigure[$\beta_{kj}=0.8$.]{
			\includegraphics[width=.32\textwidth]{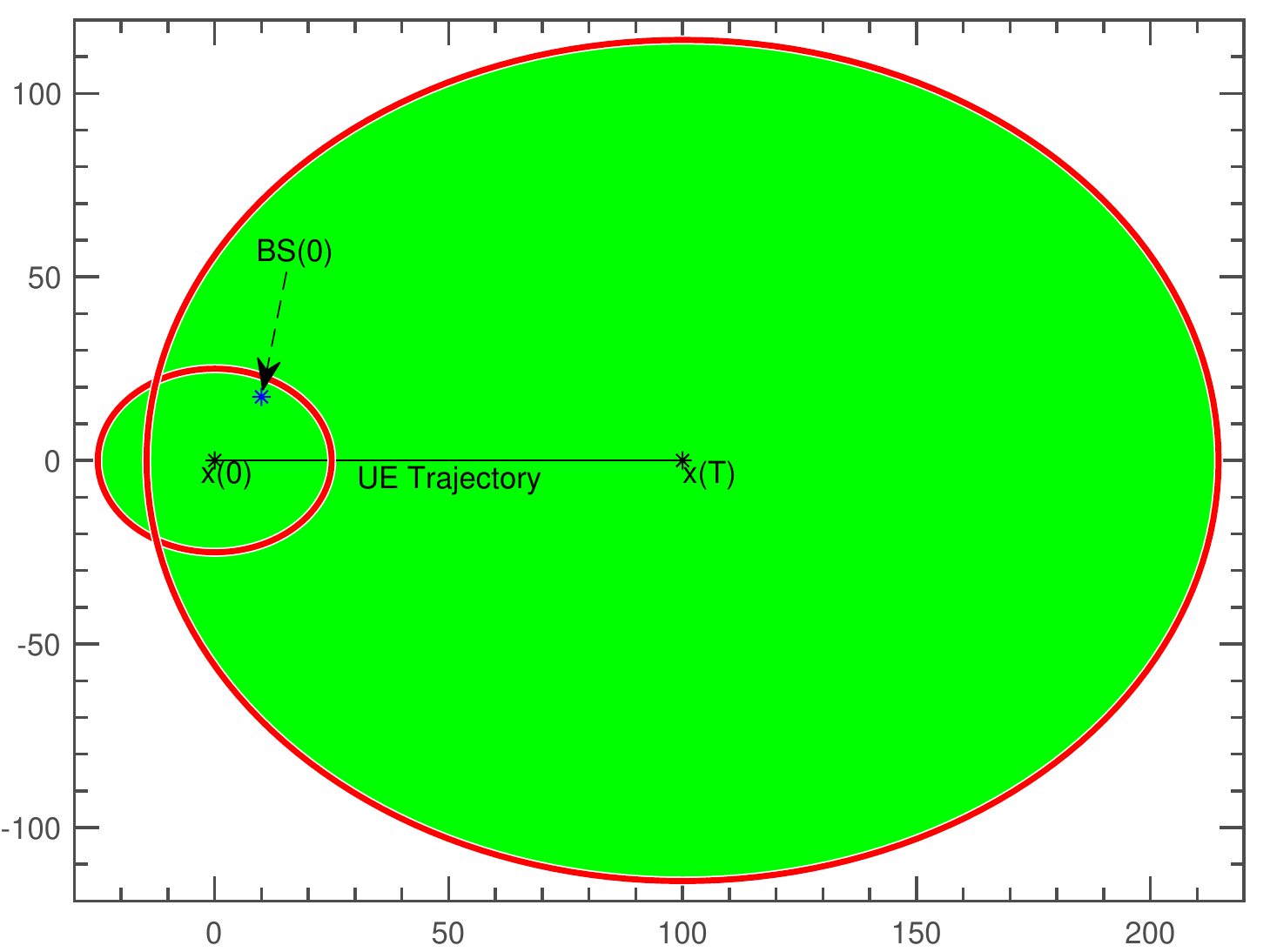}}}}
	\parbox[c]{.32\textwidth}{%
		\centerline{\subfigure[$\beta_{kj}=1$.]{
			\includegraphics[width=.32\textwidth]{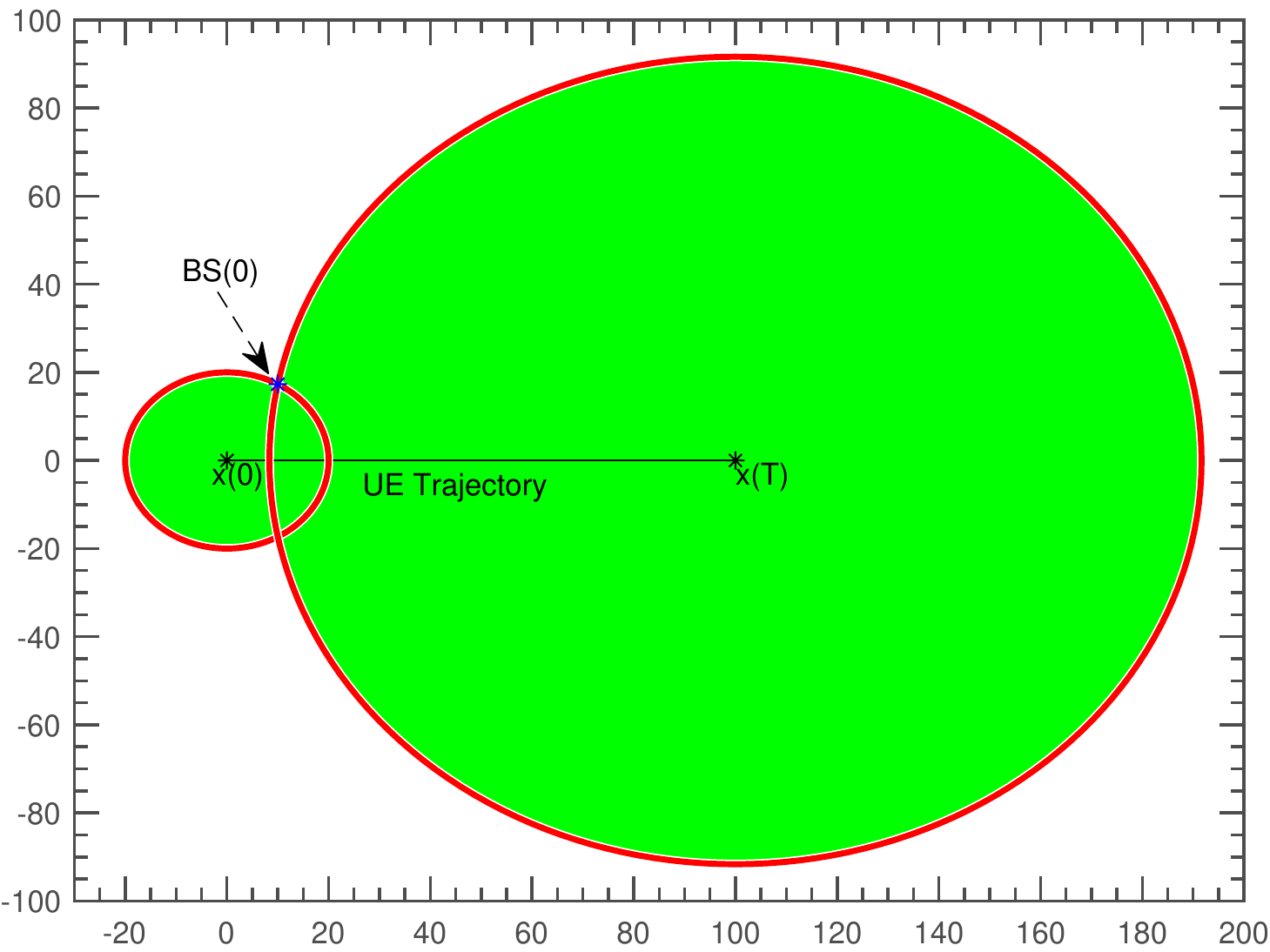}}}} 
	\parbox[c]{.32\textwidth}{%
		\centerline{\subfigure[$\beta_{kj}=1.2$.]{
			\includegraphics[width=.32\textwidth]{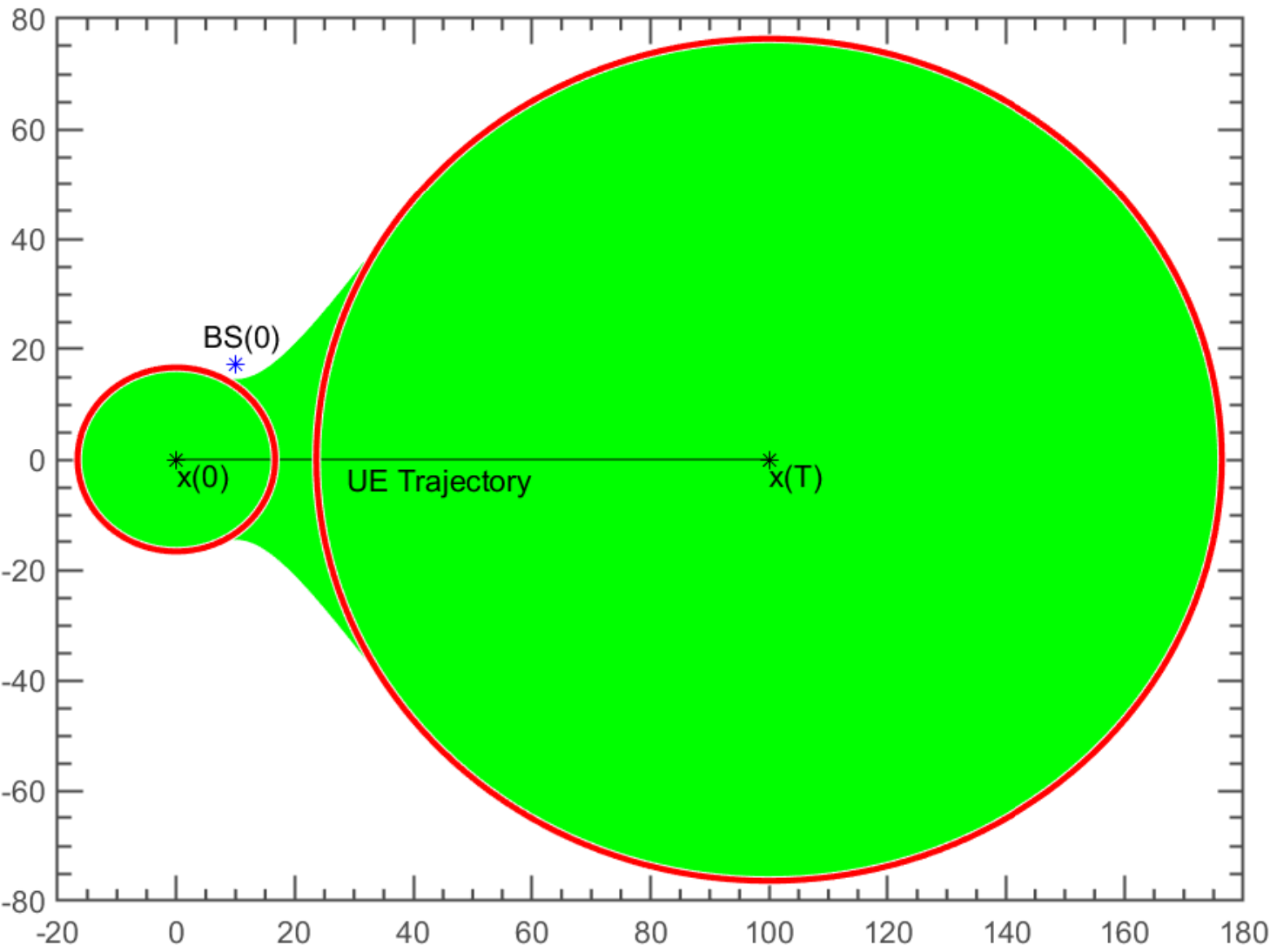}}}} 
	\caption{$\mathcal{A}_{kj}(20,\pi/3,5,20,\beta_{kj})$. (a) $\beta_{kj}<1$, (b) $\beta_{kj}=1$, and (c) $\beta_{kj}>1$. Red circles correspond to $\mathcal{B}\left( 0,\frac{r_0}{\beta_{kj}} \right)$ and $\mathcal{B}\left( \text{x}(T),\frac{r_0(T)}{\beta_{kj}} \right)$. }
	\label{fig:green_area}
\end{figure}

Using the association strategy \eqref{eq:association-policy}, we get
\begin{IEEEeqnarray}{rCl}
	\IEEEeqnarraymulticol{3}{l}{
	\mathbb{P}\left( BS(t)=BS(0), \forall t\in(0,T] \mid r_0, \theta, \text{tier}=k \right)  
	} \nonumber 
	\\
	&=& \mathbb{P}\left( \bigcap_{j\in\mathcal{K}} \Phi_j\left( \mathcal{B}\left( \text{x}(t),\frac{r_0(t)}{\beta_{kj}} \right) \setminus \mathcal{B}\left( 0,\frac{r_0}{\beta_{kj}} \right) \right)=0, \forall t\in(0,T] \mid r_0, \theta, \text{tier}=k \right),
	\label{eq:step1-2}
\end{IEEEeqnarray}
where $\mathcal{B}\left( 0,\frac{r_0}{\beta_{kj}} \right)$ is excluded since we know there is no tier $j$ BS closer than $\frac{r_0}{\beta_{kj}}$ to the typical mobile user at time 0. Let us define
\begin{IEEEeqnarray}{rCl}
	\mathcal{A}_{kj}(r_0,\theta,v,T,\beta_{kj}) = 
	\left\{ \bigcup_t \mathcal{B}\left( \text{x}(t),\frac{r_0(t)}{\beta_{kj}} \right) \mid t\in[0,T], r_0(t)=\sqrt{r_0^2+v^2t^2-2r_0vt\cos\theta} \right\}. \IEEEeqnarraynumspace
	\label{eq:def_green_area}
\end{IEEEeqnarray}
Using $\mathcal{A}_{kj}(r_0,\theta,v,T,\beta_{kj})$, we can write
\begin{IEEEeqnarray}{rCl}
	\IEEEeqnarraymulticol{3}{l}{
	\mathbb{P}\left( BS(t)=BS(0), \forall t\in(0,T] \mid r_0, \theta, \text{tier}=k \right)  
	} \nonumber
	\\
	&=& \mathbb{P}\left( \bigcap_{j\in\mathcal{K}} \Phi_j\left( \mathcal{A}_{kj}(r_0,\theta,v,T,\beta_{kj}) \setminus \mathcal{B}\left( 0,\frac{r_0}{\beta_{kj}} \right) \right)=0 \mid r_0, \theta, \text{tier}=k \right) \nonumber 
	\\
	&\stackrel{ ({\text a}) }{=}& \prod_{j\in\mathcal{K}} \mathbb{P}\left( \Phi_j\left( \mathcal{A}_{kj}(r_0,\theta,v,T,\beta_{kj}) \setminus \mathcal{B}\left( 0,\frac{r_0}{\beta_{kj}} \right) \right)=0 \mid r_0, \theta, \text{tier}=k \right) \nonumber 
	\\
	&\stackrel{ ({\text b}) }{=}& \prod_{j\in\mathcal{K}} \exp\left\{ -\lambda_{j} \left|\mathcal{A}_{kj}(r_0,\theta,v,T,\beta_{kj}) \setminus \mathcal{B}\left( 0,\frac{r_0}{\beta_{kj}} \right)\right| \right\},
	\label{eq:step1-3}
\end{IEEEeqnarray}
where $|A|$ denotes the area of $A$, (a) follows from the independence of different tiers' point processes, and (b) is obtained by using the void probability of PPP. In \figref{fig:green_area}, $\mathcal{A}_{kj}(r_0,\theta,v,T,\beta_{kj})$ is illustrated for three different cases: a) $\beta_{kj}<1$, b) $\beta_{kj}=1$, and c) $\beta_{kj}>1$. To derive the distribution of $\tilde{S}$, we need to calculate the area of $\mathcal{A}_{kj}(r_0,\theta,v,T,\beta_{kj})$ for all three cases. Further discussion about $\mathcal{A}_{kj}(r_0,\theta,v,T,\beta_{kj})$ is provided in the next section. 

\subsection{Step 2 of Analysis}

Given that, at time 0, the mobile user is associated to a tier-$k$ BS, the origin is almost surely contained in the interior of a tier-$k$ Voronoi cell. In this paper, we define \textit{linear contact distribution function} as the probability that a line segment $\ell$ containing the origin with length $r$ and random orientation crosses the cell boundaries. Therefore, given origin $o$ is inside a tier-$k$ cell, linear contact distribution function $H_{\ell}( z\mid \text{tier}=k )$ is equal to the probability that intersection of user's trajectory with length $z$ and the cell boundaries is nonempty. Using the conditional CCDF of $\tilde{S}$, we can derive the linear contact distribution function as
\begin{IEEEeqnarray}{rCl}
	H_{\ell}( z\mid \text{tier}=k ) 
	&=& 1 -\mathbb{P}( \tilde{S}>\frac{z}{v} \mid \text{tier} = k ) = 1 - \bar{F}_{\tilde{S}}( \frac{z}{v} \mid \text{tier} = k ).
	\label{eq:def_contact_distribution_function}
\end{IEEEeqnarray} 

\subsection{Step 3 of Analysis}
So far, we have considered the sojourn time in the cell where connection is initiated ($\tilde{S}$). Distribution of the sojourn time ($S$), for tier-$k$, can be obtained using the chord length distribution. Due to the stationarity of our model, chord length distribution for tier-$k$, denoted by $F_{L}(z\mid\text{tier}=k)$, can be computed as follows \cite{chiu2013stochastic}:
\begin{IEEEeqnarray}{rCl}
	F_{L}(z\mid\text{tier}=k) = 1-\mathbb{E}[L \mid \text{tier}=k] \frac{{\rm d}}{{\rm d}z} H_{\ell}( z\mid \text{tier}=k ),
	\label{eq:def_chord_length_distribution_function}
\end{IEEEeqnarray}
where $\mathbb{E}[L \mid \text{tier}=k]$ is the mean length of the chords lying in tier-$k$ cells, and is obtained by \cite{heinrich1998contact}
\begin{IEEEeqnarray}{rCl}
	\mathbb{E}[L \mid \text{tier}=k] = \lim_{z \to 0} \frac{z}{ H_{\ell}( z\mid \text{tier}=k ) }
	\label{eq:def_mean_chord_length}
\end{IEEEeqnarray}

\subsection{Step 4 of Analysis} 
Finally, we can characterize the sojourn time distribution for tier-$k$ using the results from previous step. In particular, the mean and CCDF of the sojourn time in tier-$k$ are 
\begin{IEEEeqnarray}{rCl}
	\mathbb{E}[S \mid \text{tier}=k ] &=& \frac{1}{v} \mathbb{E}[L \mid \text{tier}=k], \label{eq:def_mean_sojourn_time}  \\
	\bar{F}_S(T \mid \text{tier}=k) &=& 1-F_{L}(vT\mid\text{tier}=k). \label{eq:def_distribution_sojourn_time}  
\end{IEEEeqnarray}

\section{First Step of Sojourn Time Analysis: Derivation of $|\mathcal{A}_{kj}(r_0,\theta,v,T,\beta_{kj})|$}

As mentioned in the previous subsection, the first step of sojourn time analysis requires calculation of $|\mathcal{A}_{kj}(r_0,\theta,v,T,\beta_{kj})|$ (area of $\mathcal{A}_{kj}(r_0,\theta,v,T,\beta_{kj})$). In this regard, we consider three cases: I) $\beta_{kj}<1$, II) $\beta_{kj}>1$, and III) $\beta_{kj}=1$.   


\subsection{Case I: $\beta_{kj}<1$}

The following proposition helps us to derive the area of $\mathcal{A}_{kj}(r_0,\theta,v,T,\beta_{kj})$ for this case.
\begin{Proposition} \label{Prop1}
	When $\beta_{kj}<1$, $\mathcal{A}_{kj}(r_0,\theta,v,T,\beta_{kj}) = \mathcal{B}\left( \textup{x}(0),\frac{r_0}{\beta_{kj}} \right) \cup \mathcal{B}\left( \textup{x}(T),\frac{r_0(T)}{\beta_{kj}} \right) $.
\end{Proposition}
\begin{IEEEproof}
	See \textbf{Appendix A}.
\end{IEEEproof}
Note that, depending on radii of the two circles, $\frac{r_0}{\beta_{kj}}$ and $\frac{r_0(T)}{\beta_{kj}}$, and the distance between their centres, i.e. $vT$,
three different situations can happen when $\beta_{kj}<1$:

\vspace{0.2cm}
{\em Situation 1}: When $\frac{r_0(T)}{\beta_{kj}} \ge \frac{r_0}{\beta_{kj}} + vT$, we have $\mathcal{B}\left( \textup{x}(0),\frac{r_0}{\beta_{kj}} \right) \subset \mathcal{B}\left( \textup{x}(T),\frac{r_0(T)}{\beta_{kj}} \right)$, which yields $\mathcal{A}_{kj}(r_0,\theta,v,T,\beta_{kj}) = \mathcal{B}\left( \textup{x}(T),\frac{r_0(T)}{\beta_{kj}} \right)$.

\vspace{0.2cm}
{\em Situation 2}: When $\frac{r_0}{\beta_{kj}} \ge \frac{r_0(T)}{\beta_{kj}} + vT$, we have $\mathcal{B}\left( \textup{x}(T),\frac{r_0(T)}{\beta_{kj}} \right) \subset \mathcal{B}\left( \textup{x}(0),\frac{r_0}{\beta_{kj}} \right)$, which yields $\mathcal{A}_{kj}(r_0,\theta,v,T,\beta_{kj}) = \mathcal{B}\left( \textup{x}(0),\frac{r_0}{\beta_{kj}} \right) $.

\vspace{0.2cm}
{\em Situation 3}: When $\frac{r_0(T)}{\beta_{kj}} < \frac{r_0}{\beta_{kj}} + vT$ and $\frac{r_0}{\beta_{kj}} < \frac{r_0(T)}{\beta_{kj}} + vT$, $\mathcal{A}_{kj}(r_0,\theta,v,T,\beta_{kj}) = \mathcal{B}\left( \textup{x}(0),\frac{r_0}{\beta_{kj}} \right) \cup \mathcal{B}\left( \textup{x}(T),\frac{r_0(T)}{\beta_{kj}} \right) $. An example of which is illustrated in \figref{fig:green_area}(a). 

\vspace{0.2cm}
Using this information, now we can compute $|\mathcal{A}_{kj}(r_0,\theta,v,T,\beta_{kj})|$ when $\beta_{kj}<1$.
\begin{IEEEeqnarray}{rCl}
		|\mathcal{A}_{kj}(r_0,\theta,v,T,\beta_{kj})| =  
	\begin{cases}
		\pi \frac{r_0(T)^2}{\beta_{kj}^2}, & \text{if } 2 r_0  \frac{ cos\theta+\beta_{kj} }{ 1-\beta_{kj}^2 } \le vT \\
		\pi \frac{r_0^2}{\beta_{kj}^2},    & \text{if } vT \le 2 r_0  \frac{ cos\theta-\beta_{kj} }{ 1-\beta_{kj}^2 } \\
		\pi \frac{r_0^2}{\beta_{kj}^2} + \pi\frac{r_0(T)^2}{\beta_{kj}^2} - V\left( \frac{r_0}{\beta_{kj}},\frac{r_0(T)}{\beta_{kj}},vT \right), & 
		\text{if } 2 r_0  \frac{ cos\theta-\beta_{kj} }{ 1-\beta_{kj}^2 } < vT < 2 r_0  \frac{ cos\theta+\beta_{kj} }{ 1-\beta_{kj}^2 }
	\end{cases}
	\label{eq:area_beta_lt_1}
\end{IEEEeqnarray}
where $V\left( \frac{r_0}{\beta_{kj}},\frac{r_0(T)}{\beta_{kj}},vT \right)$ is the area of intersection of two circles with radii $\frac{r_0}{\beta_{kj}}$ and $\frac{r_0(T)}{\beta_{kj}}$ whose centers are separated by $vT$, i.e., $V\left( \frac{r_0}{\beta_{kj}},\frac{r_0(T)}{\beta_{kj}},vT \right) =$
\begin{IEEEeqnarray}{rCl} 
	    && \frac{r_0^2}{\beta_{kj}^2} \arccos\left( \frac{r_0^2+\beta_{kj}^2v^2T^2-r_0(T)^2}{2\beta_{kj}r_0vT} \right) + \frac{r_0(T)^2}{\beta_{kj}^2} \arccos\left( \frac{r_0(T)^2+\beta_{kj}^2v^2T^2-r_0^2}{2\beta_{kj}r_0(T)vT} \right) \nonumber 
	    \\
	    && - \frac{1}{2} \sqrt{ \left( \frac{r_0}{\beta_{kj}} + \frac{r_0(T)}{\beta_{kj}} + vT \right)\left( \frac{r_0}{\beta_{kj}}+\frac{r_0(T)}{\beta_{kj}} - vT \right)\left( \frac{r_0}{\beta_{kj}} - \frac{r_0(T)}{\beta_{kj}} + vT \right)\left( -\frac{r_0}{\beta_{kj}}+\frac{r_0(T)}{\beta_{kj}} + vT \right) }.
	    \nonumber \\
	    \label{eq:area_of_intersection}
\end{IEEEeqnarray}

\begin{figure}
	\parbox[c]{.5\textwidth}{%
		\centerline{\subfigure[$t=2,2.2$.]{
			\includegraphics[width=.5\textwidth]{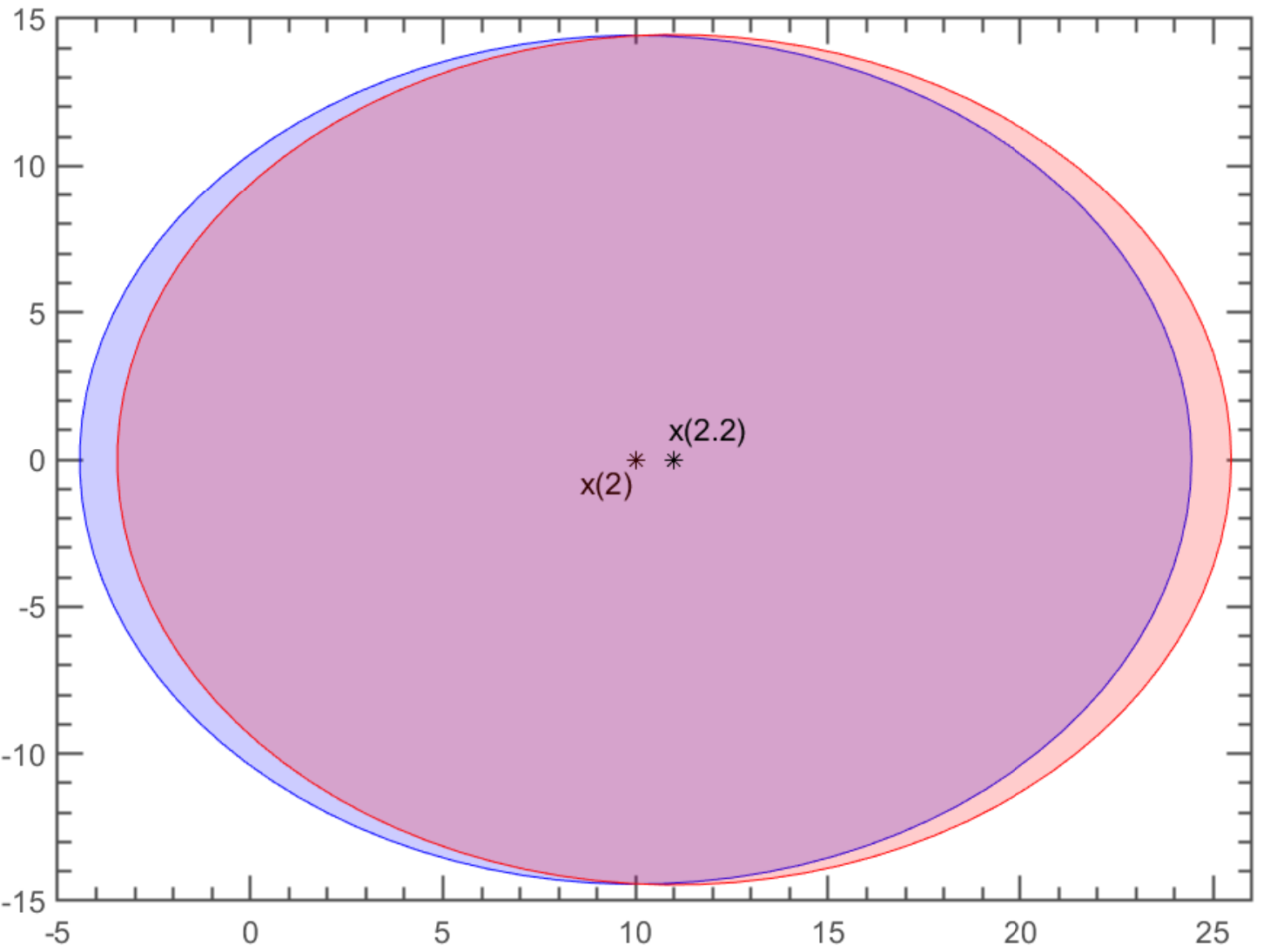}}}}
	\parbox[c]{.5\textwidth}{%
		\centerline{\subfigure[$t=0,1,2,3,4,5,6,7,8$.]{
			\includegraphics[width=.5\textwidth]{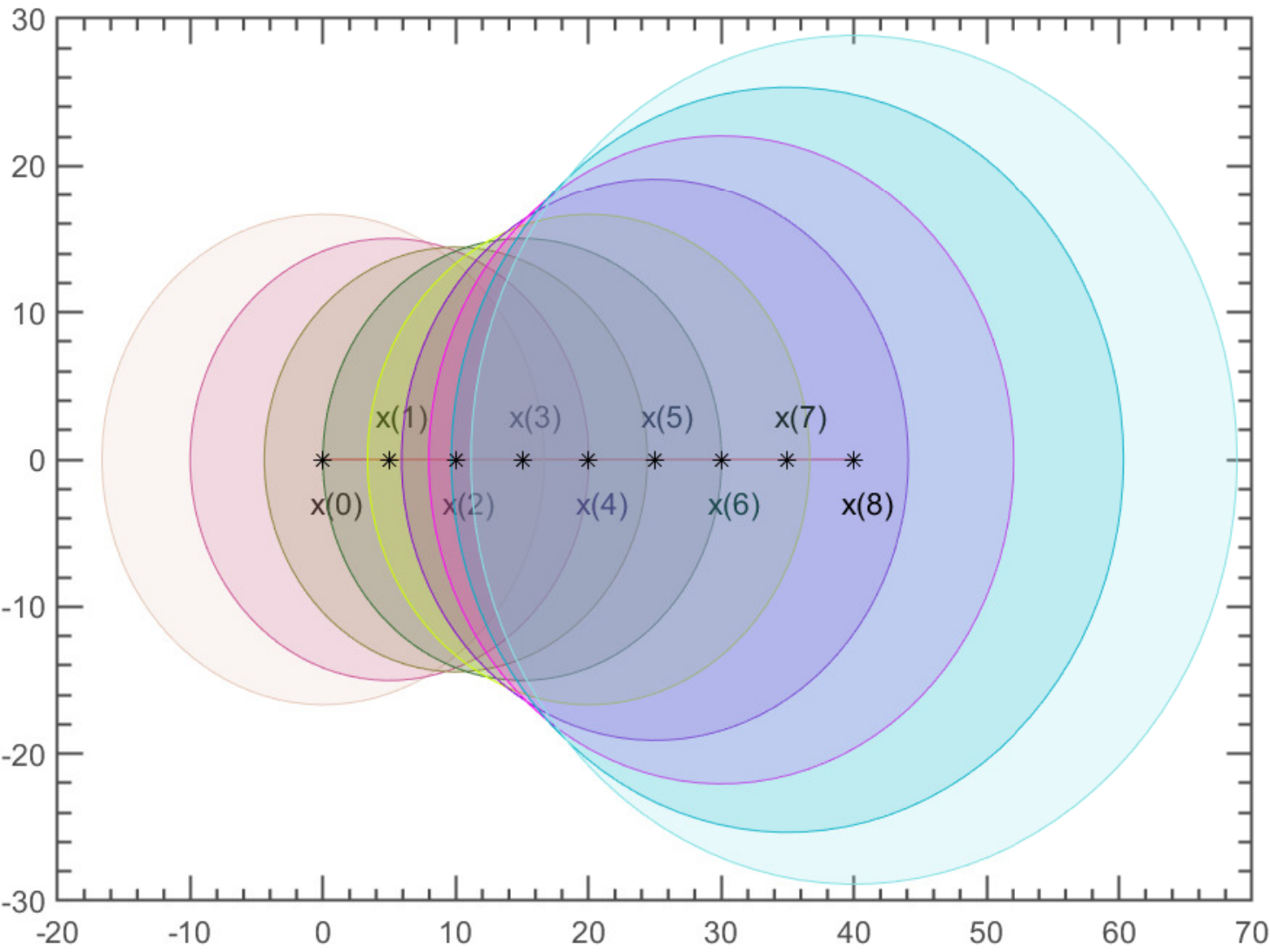}}}}	
	\caption{$\mathcal{B}\left( \textup{x}(t),\frac{r_0(t)}{\beta_{kj}} \right)$ at different time instants. Union of these circles from $t=0$ till $t=8$ forms $\mathcal{A}_{kj}(20,\pi/3,5,8,1.2)$.}
	\label{fig:beta_gt_1}
\end{figure}


\subsection{Case II: $B_{kj}>1$}

For this case, to derive the area of $\mathcal{A}_{kj}(r_0,\theta,v,T,\beta_{kj})$,  first we study the intersection of $\mathcal{B}\left( \textup{x}(t),\frac{r_0(t)}{\beta_{kj}} \right)$ and $\mathcal{B}\left( \textup{x}(t+{\rm d}t),\frac{r_0(t+{\rm d}t)}{\beta_{kj}} \right)$ as ${\rm d}t\to0$ (\figref{fig:beta_gt_1}(a)). From triangle equations, we have 
\begin{IEEEeqnarray}{rCl}
	r_0(t+{\rm d}t)^2=r_0^2+v^2(t+{\rm d}t)^2-2r_0v(t+{\rm d}t)\cos\theta=r_0(t)^2+v^2{\rm d}t^2+2v{\rm d}t(vt-r_0\cos\theta). \IEEEeqnarraynumspace
	\label{eq:r_t_dt}
\end{IEEEeqnarray}
Since $|vt-r_0\cos\theta| \le r_0(t)$, 
\begin{IEEEeqnarray}{rCl}
	r_0(t)^2+v^2{\rm d}t^2-2r_0(t)v{\rm d}t \le r_0(t+{\rm d}t)^2 \le r_0(t)^2+v^2{\rm d}t^2+2r_0(t)v{\rm d}t . 
	\label{eq:r_inequality} 
\end{IEEEeqnarray}
Dividing  $r_0(t) - v{\rm d}t \le r_0(t+{\rm d}t) \le r_0(t) + v{\rm d}t$ by $\beta_{kj}$ yields
\begin{IEEEeqnarray}{rCl}
	\frac{r_0(t)}{\beta_{kj}} - v{\rm d}t \stackrel{(\text{a})}{\le} \frac{r_0(t)}{\beta_{kj}} - \frac{v{\rm d}t}{\beta_{kj}} \le
	\frac{r_0(t+{\rm d}t)}{\beta_{kj}} \le \frac{r_0(t)}{\beta_{kj}} + \frac{v{\rm d}t}{\beta_{kj}} \stackrel{(\text{b})}{\le} \frac{r_0(t)}{\beta_{kj}} + v{\rm d}t, \nonumber
\end{IEEEeqnarray}
where (a) and (b) are obtained using $\beta_{kj}>1$. Therefore, as ${\rm d}t\to0$, $\mathcal{B}\left( \textup{x}(t),\frac{r_0(t)}{\beta_{kj}} \right)$ and $\mathcal{B}\left( \textup{x}(t+{\rm d}t),\frac{r_0(t+{\rm d}t)}{\beta_{kj}} \right)$ partially overlap (boundaries of $\mathcal{B}\left( \textup{x}(t),\frac{r_0(t)}{\beta_{kj}} \right)$ and $\mathcal{B}\left( \textup{x}(t+{\rm d}t),\frac{r_0(t+{\rm d}t)}{\beta_{kj}} \right)$ intersect at two points), and we have 
\begin{IEEEeqnarray}{rCl}
	\IEEEeqnarraymulticol{3}{l}{
	\left| \mathcal{B}\left( \textup{x}(t),\frac{r_0(t)}{\beta_{kj}} \right) \setminus \mathcal{B}\left( \textup{x}(t+{\rm d}t),\frac{r_0(t+{\rm d}t)}{\beta_{kj}} \right) \right| = \pi \frac{r_0(t)^2}{\beta_{kj}^2} - V\left(\frac{r_0(t)}{\beta_{kj}},\frac{r_0(t+{\rm d}t)}{\beta_{kj}},v{\rm d}t\right) } \nonumber 
	\\ 
	&=& \pi \frac{r_0(t)^2}{\beta_{kj}^2} - \arccos\left( \frac{r_0\cos\theta-vt}{\beta_{kj}r_0(t)} + \frac{\beta_{kj}^2-1}{2\beta_{kj}} \frac{v{\rm d}t}{r_0(t)} \right) \frac{r_0(t)^2}{\beta_{kj}^2} \nonumber \\
	&&- \arccos\left( \frac{vt-r_0\cos\theta}{\beta_{kj}r_0(t+{\rm d}t)} + \frac{\beta_{kj}^2+1}{2\beta_{kj}} \frac{v{\rm d}t}{r_0(t+{\rm d}t)} \right)
	\frac{r_0(t+{\rm d}t)^2}{\beta_{kj}^2} \nonumber \\
	&&+\frac{1}{2}\sqrt{ \frac{2v{\rm d}t}{\beta_{kj}}\left( r_0(t)-\frac{vt-r_0\cos\theta}{\beta_{kj}} \right) + v^2{\rm d}t^2 \left( 1-\frac{1}{\beta_{kj}^2} \right) } \nonumber \\
	&&\times \sqrt{ \frac{2v{\rm d}t}{\beta_{kj}}\left( r_0(t)+\frac{vt-r_0\cos\theta}{\beta_{kj}} \right) - v^2{\rm d}t^2 \left( 1-\frac{1}{\beta_{kj}^2} \right) } \nonumber \\
	&=& \frac{2v}{\beta_{kj}^2} \left[\sqrt{\beta_{kj}^2r_0(t)^2-(vt-r_0\cos\theta)^2}-
	    \arccos\left( \frac{vt-r_0\cos\theta}{\beta_{kj}r_0(t)} \right)(vt-r_0\cos\theta) \right]{\rm d}t + O({\rm d}t^2), \nonumber \\
	\label{eq:infinitesimal}
\end{IEEEeqnarray}
where the last result is proved in \textbf{Appendix~B}.

Moreover, we can derive the intersection points of these two circles from their equations in Cartesian coordinate system, i.e.,
\begin{IEEEeqnarray}{rCl}
	\mathcal{B}\left( \textup{x}(t),\frac{r_0(t)}{\beta_{kj}} \right) &:& \left[x-vt\right]^2+y^2=\frac{r_0(t)^2}{\beta_{kj}^2}, \nonumber \\
	\mathcal{B}\left( \textup{x}(t+{\rm d}t),\frac{r_0(t+{\rm d}t)}{\beta_{kj}} \right) &:& \left[x-v(t+{\rm d}t)\right]^2+y^2=\frac{r_0(t+{\rm d}t)^2}{\beta_{kj}^2}. \nonumber
\end{IEEEeqnarray}
Combining these equations and solving for $x$ results in
\begin{IEEEeqnarray}{rCl}
	x = v\left(t+\frac{{\rm d}t}{2}\right)\left( 1-\frac{1}{\beta_{kj}^2} \right)+\frac{r_0\cos\theta}{\beta_{kj}^2}, \nonumber
\end{IEEEeqnarray}
which indicates that, for $\beta_{kj}>1$, the boundaries' intersection points move along the positive $x$-axis as $t$ increases (\figref{fig:beta_gt_1}(b)). Using this result, we can write
\begin{IEEEeqnarray}{rCl}
	\IEEEeqnarraymulticol{3}{l}{
	\left| \bigcup_{i=0}^n \mathcal{B}\left( \textup{x}(t+i{\rm d}t),\frac{r_0(t+i{\rm d}t)}{\beta_{kj}} \right) \right| =
	\pi \frac{r_0(t+n{\rm d}t)^2}{\beta_{kj}^2}} \nonumber \\
	&&+ \sum_{i=0}^{n-1} \left| \mathcal{B}\left( \textup{x}(t+i{\rm d}t),\frac{r_0(t+i{\rm d}t)}{\beta_{kj}} \right)  \setminus \mathcal{B}\left( \textup{x}(t+(i+1){\rm d}t),\frac{r_0(t+(i+1){\rm d}t)}{\beta_{kj}} \right) \right| 
	\label{eq:beta_gt_1_basic}
\end{IEEEeqnarray} 

Let the time interval $[0,T]$ be partitioned by points $t_i=i{\rm d}t$, $i=0,...,\frac{T}{{\rm d}t}$.
We can calculate the area of $\mathcal{A}_{kj}(r_0,\theta,v,T,\beta_{kj})$ by setting $t=0$ in \eqref{eq:beta_gt_1_basic}, i.e.,
\begin{IEEEeqnarray}{rCl}
 	\IEEEeqnarraymulticol{3}{l}{
 	|\mathcal{A}_{kj}(r_0,\theta,v,T,\beta_{kj})| = \lim_{{\rm d}t\to0} \left| \bigcup_{i=0}^\frac{T}{{\rm d}t} \mathcal{B}\left( \textup{x}(i{\rm d}t),\frac{r_0(i{\rm d}t)}{\beta_{kj}} \right) \right|} \nonumber 
 	\\
 	&=& \pi \frac{r_0(T)^2}{\beta_{kj}^2} + \lim_{{\rm d}t\to0} \sum_{i=0}^{\frac{T}{{\rm d}t}-1} \left| \mathcal{B}\left( \textup{x}(i{\rm d}t),\frac{r_0(i{\rm d}t)}{\beta_{kj}} \right)  \setminus \mathcal{B}\left( \textup{x}((i+1){\rm d}t),\frac{r_0((i+1){\rm d}t)}{\beta_{kj}} \right) \right|  \nonumber 
 	\\
 	&\stackrel{\text{(a)}}{=}& \pi \frac{r_0(T)^2}{\beta_{kj}^2} + \lim_{{\rm d}t\to0} \sum_{i=0}^{\frac{T}{{\rm d}t}-1} \nonumber 
 	\\
 	&&  \frac{2v}{\beta_{kj}^2} \left[\sqrt{\beta_{kj}^2r_0(t_i)^2-(vt_i-r_0\cos\theta)^2}-
 	\arccos\left( \frac{vt_i-r_0\cos\theta}{\beta_{kj}r_0(t_i)} \right)(vt_i-r_0\cos\theta) \right]{\rm d}t + O({\rm d}t^2) \nonumber 
 	\\
 	&\stackrel{\text{(b)}}{=}& \pi \frac{r_0(T)^2}{\beta_{kj}^2} + \frac{2v}{\beta_{kj}^2} \int_0^T \sqrt{\beta_{kj}^2r_0(t)^2-(vt-r_0\cos\theta)^2}-
 	\arccos\left( \frac{vt-r_0\cos\theta}{\beta_{kj}r_0(t)} \right)(vt-r_0\cos\theta) {\rm d}t, \nonumber \\
 	\label{eq:area_beta_gt_1}
\end{IEEEeqnarray}
where (a) is obtained by using \eqref{eq:infinitesimal} and (b) follows from the Riemann integral.


\subsection{Case III: $\beta_{kj}=1$}

For this case, similar to \textbf{Appendix~A}, we can prove $\mathcal{A}_{kj}(r_0,\theta,v,T,1) = \mathcal{B}\left( \textup{x}(0),r_0 \right) \cup \mathcal{B}\left( \textup{x}(T),r_0(T) \right)$. Since $|r_0-vT| \le r_0(T) \le r_0+vT$, $\mathcal{B}\left( \textup{x}(0),r_0 \right)$ and $\mathcal{B}\left( \textup{x}(T),r_0(T) \right)$ partially overlap. Therefore,
\begin{IEEEeqnarray}{rCl}
	|\mathcal{A}_{kj}(r_0,\theta,v,T,\beta_{kj})| =
	\pi \frac{r_0^2}{\beta_{kj}^2} + \pi\frac{r_0(T)^2}{\beta_{kj}^2} - V\left( \frac{r_0}{\beta_{kj}},\frac{r_0(T)}{\beta_{kj}},vT \right),
	\label{eq:area_beta_1}
\end{IEEEeqnarray}
where $\beta_{kj}=1$ and $V\left( \frac{r_0}{\beta_{kj}},\frac{r_0(T)}{\beta_{kj}},vT \right)$ is given in \eqref{eq:area_of_intersection}.

\subsection{Closed-form Expression for $|\mathcal{A}_{kj}(r_0,\theta,v,T,\beta_{kj})|$}

\begin{Theorem}\label{Thm1}
Area of $\mathcal{A}_{kj}(r_0,\theta,v,T,\beta_{kj})$ can be obtained by	
\begin{IEEEeqnarray}{rCl}
	\IEEEeqnarraymulticol{3}{l}{
	|\mathcal{A}_{kj}(r_0,\theta,v,T,\beta_{kj})| = |\mathcal{A}_{kj}(r_0,\theta,vT,1,\beta_{kj})| = } \nonumber \\
	\begin{cases}
		\pi \frac{g(vT,1)^2}{\beta_{kj}^2} + \frac{2vT}{\beta_{kj}^2} \int\limits_0^1 \sqrt{\beta_{kj}^2g(vT,u)^2-(vTu-r_0\cos\theta)^2}-
		\arccos\left( \frac{vTu-r_0\cos\theta}{\beta_{kj}g(vT,u)} \right)(vTu-r_0\cos\theta) {\rm d}u, 
		\\
		\qquad \qquad \qquad \qquad \qquad \qquad \qquad \qquad \qquad \qquad \qquad \qquad \qquad \qquad \qquad \qquad \quad \! \!
		\text{if } \left( \beta_{kj}>1 \right) 
		\\
		\pi \frac{r_0^2}{\beta_{kj}^2} + \pi\frac{g(vT,1)^2}{\beta_{kj}^2} - V\left( \frac{r_0}{\beta_{kj}},\frac{g(vT,1)}{\beta_{kj}},vT \right), 
		\\
		\qquad \qquad \qquad \qquad \qquad \qquad \! \text{if } \left( \beta_{kj}=1 \right) \text{ or } \left( \beta_{kj}<1 \text{ and } 2 r_0  \frac{ cos\theta-\beta_{kj} }{ 1-\beta_{kj}^2 } < vT < 2 r_0  \frac{ cos\theta+\beta_{kj} }{ 1-\beta_{kj}^2 } \right) 
		\\
		\pi \frac{r_0^2}{\beta_{kj}^2}, 
		\qquad \qquad \qquad \qquad \qquad \qquad \qquad \qquad \qquad \qquad \quad
		\text{if } \left( \beta_{kj}<1 \text{ and } vT \le 2 r_0  \frac{ cos\theta-\beta_{kj} }{ 1-\beta_{kj}^2 } \right) 
		\\
		\pi \frac{g(vT,1)^2}{\beta_{kj}^2}, 
		\qquad \qquad \qquad \qquad \qquad \qquad \qquad \qquad \qquad \qquad \! \! \!
		 \text{if } \left( \beta_{kj}<1 \text{ and } 2 r_0  \frac{ cos\theta+\beta_{kj} }{ 1-\beta_{kj}^2 } \le vT \right)
	\end{cases},\nonumber \\ 
	\label{eq:A}
\end{IEEEeqnarray}
where $g(vT,u)=r_0(Tu)=\sqrt{r_0^2+v^2T^2u^2-2r_0vTu\cos\theta}$ and $V\left( \frac{r_0}{\beta_{kj}},\frac{g(vT,1)}{\beta_{kj}},vT \right)$ is given in \eqref{eq:area_of_intersection}.
\end{Theorem}
\begin{IEEEproof}
	The proof follows from combining \eqref{eq:area_beta_lt_1}, \eqref{eq:area_beta_gt_1}, and \eqref{eq:area_beta_1}. For, $\beta_{kj}>1$, we have used change of variable $\frac{t}{T}=u$.
\end{IEEEproof}

It is worth mentioning that, according to \textbf{Theorem~1}, in multi-tier networks, tier-$k$ cells are convex,  if $B_kP_k \le B_jP_j$ (or equivalently, $\beta_{kj}\le1$), $\forall j\in\mathcal{K}$. Therefore, in this case, we can derive the sojourn time distribution of tier-$k$ similar to the single-tier scenario using \eqref{eq-sojourn-time-def-single-tier} (instead of \eqref{eq-sojourn-time-def}). This is the reason why some works in the literature only focus on the sojourn time in small cells of two-tier networks.

\section{Main Results on Sojourn Time and Handoff Rate and Effects of Network Parameters}

\subsection{Conditional CCDF and Mean of Sojourn Time}

Since $\mathcal{B}\left( 0,\frac{r_0}{\beta_{kj}} \right) \subset \mathcal{A}_{kj}(r_0,\theta,v,T,\beta_{kj})$, we can further simplify \eqref{eq:step1-3} as
\begin{multline}
	\mathbb{P}\left( BS(t)=BS(0), \forall t\in(0,T] \mid r_0, \theta, \text{tier}=k \right)  
	= \\
	\exp\left\{ -\sum_{j\in\mathcal{K}} \lambda_{j} \left( \left|\mathcal{A}_{kj}(r_0,\theta,vT,1,\beta_{kj})\right| - \pi\frac{r_0^2}{\beta_{kj}^2} \right) \right\}, 
	\label{eq:Stilde_basic}
\end{multline}
The CCDF of the sojourn time of a connection in a cell where it is initiated, $\tilde{S}$ can be obtained by substituting \eqref{eq:PDF_r0} and \eqref{eq:Stilde_basic} in \eqref{eq:Step1-1}.
\begin{IEEEeqnarray}{rCl}
	\IEEEeqnarraymulticol{3}{l}{
	\bar{F}_{\tilde{S}}(T \mid \text{tier}=k) =} \nonumber 
	\\
	&& \frac{1}{\mathbb{P}(\text{tier}=k)} \int_0^{\infty} \int_0^{\pi} 2\lambda_k r_0 \exp\left\{ -\sum_{j\in\mathcal{K}} \lambda_{j} \left|\mathcal{A}_{kj}(r_0,\theta,vT,1,\beta_{kj})\right| \right\} {\rm d}\theta {\rm d}r_0, 
	\label{eq:Stilde_conditional}
\end{IEEEeqnarray}
where $\mathbb{P}(\text{tier}=k)$ is given in \eqref{eq:tier_k}, and $\left|\mathcal{A}_{kj}(r_0,\theta,vT,1,\beta_{kj})\right|$ is given in \textbf{Theorem~1}.

As discussed before, in Step 2, we use \eqref{eq:Stilde_conditional} to derive the linear contact distribution function given that the mobile user is in a tier-$k$ cell at time 0, i.e.
\begin{IEEEeqnarray}{rCl}
	\IEEEeqnarraymulticol{3}{l}{
	H_{\ell}( z\mid \text{tier}=k ) 
	= 1 - \bar{F}_{\tilde{S}}( \frac{z}{v} \mid \text{tier} = k ) = } \nonumber \\
	&& 1 - \frac{1}{\mathbb{P}(\text{tier}=k)} \int_0^{\infty} \int_0^{\pi} 2\lambda_k r_0 \exp\left\{ -\sum_{j\in\mathcal{K}} \lambda_{j} \left|\mathcal{A}_{kj}(r_0,\theta,z,1,\beta_{kj})\right| \right\} {\rm d}\theta {\rm d}r_0.
	\label{eq:contact_distribution}
\end{IEEEeqnarray} 

To derive the chord length distribution in tier-$k$ cells, according to \eqref{eq:def_chord_length_distribution_function}, we need $\frac{{\rm d}}{{\rm d}z} H_{\ell}( z\mid \text{tier}=k )$ and $\mathbb{E}[L \mid \text{tier}=k]$. From \eqref{eq:contact_distribution}, we have
\begin{IEEEeqnarray}{rCl}
	\frac{{\rm d}}{{\rm d}z} H_{\ell}( z\mid \text{tier}=k ) &=& 
	\frac{1}{\mathbb{P}(\text{tier}=k)} \int_0^{\infty} \int_0^{\pi} 2\lambda_k r_0 \left( \sum_{j\in\mathcal{K}} \lambda_{j} \frac{{\rm d}}{{\rm d}z}  \left|\mathcal{A}_{kj}(r_0,\theta,z,1,\beta_{kj})\right| \right)  
	\nonumber \\
	&& \times \exp\left\{ -\sum_{j\in\mathcal{K}} \lambda_{j} \left|\mathcal{A}_{kj}(r_0,\theta,z,1,\beta_{kj})\right| \right\} {\rm d}\theta {\rm d}r_0,
	\label{eq:derivative_conatct_distribution}
\end{IEEEeqnarray}
where 
\begin{IEEEeqnarray}{rCl}
	\IEEEeqnarraymulticol{3}{l}{
		\frac{{\rm d}}{{\rm d}z} |\mathcal{A}_{kj}(r_0,\theta,z,1,\beta_{kj})| = } \nonumber \\
	\begin{cases}
		\pi \frac{2(z-r_0\cos\theta)}{\beta_{kj}^2} + \frac{2}{\beta_{kj}^2} \int\limits_0^1 \sqrt{\beta_{kj}^2g(z,u)^2-(zu-r_0\cos\theta)^2}-
		\arccos\left( \frac{zu-r_0\cos\theta}{\beta_{kj}g(z,u)} \right)(zu-r_0\cos\theta) {\rm d}u 
		\\
		+ \frac{2z}{\beta_{kj}^2} \int\limits_0^1 \sqrt{\beta_{kj}^2g(z,u)^2-(zu-r_0\cos\theta)^2}\frac{u(zu-r_0\cos\theta)}{g(z,u)^2}-
		\arccos\left( \frac{zu-r_0\cos\theta}{\beta_{kj}g(z,u)} \right)u {\rm d}u, 
		\\
		\qquad \qquad \qquad \qquad \qquad \qquad \qquad \qquad \qquad \qquad \qquad \qquad \qquad \qquad \qquad \qquad \quad \! \!
		\text{if } \left( \beta_{kj}>1 \right) 
		\\
		\pi\frac{2(z-r_0\cos\theta)}{\beta_{kj}^2} + \frac{ \frac{\beta_{kj}^2-1}{\beta_{kj}^2}r_0^2 }{\sqrt{4\beta_{kj}^2r_0^2-\left(\left(\beta_{kj}^2-1\right)z+2r_0\cos\theta\right)^2}} + 
		\frac{ \frac{\beta_{kj}^2+1-2\cos^2\theta}{\beta_{kj}^2}r_0^2-\frac{\beta_{kj}^2-1}{\beta_{kj}^2}r_0z\cos\theta }{\sqrt{4\beta_{kj}^2g(z,1)^2-\left(\left(\beta_{kj}^2+1\right)z-2r_0\cos\theta\right)^2}} 
		\\
		- \arccos\left( \frac{\left(\beta_{kj}^2+1\right)z-2r_0\cos\theta}{2\beta_{kj}g(z,1)} \right) \frac{2(z-r_0\cos\theta)}{\beta_{kj}^2}
		+ \sqrt{ \frac{-\left(\beta_{kj}^2-1\right)z+2r_0\left(\beta_{kj}-\cos\theta\right)}{\left(\beta_{kj}^2-1\right)z+2r_0\left(\beta_{kj}+\cos\theta\right)} }
		\frac{\left(\beta_{kj}^2-1\right)z+r_0\left(\beta_{kj}+\cos\theta\right)}{2\beta_{kj}^2} 
		\\
		+ \sqrt{ \frac{\left(\beta_{kj}^2-1\right)z+2r_0\left(\beta_{kj}+\cos\theta\right)}{-\left(\beta_{kj}^2-1\right)z+2r_0\left(\beta_{kj}-\cos\theta\right)}}
		\frac{-\left(\beta_{kj}^2-1\right)z+r_0\left(\beta_{kj}-\cos\theta\right)}{2\beta_{kj}^2}
		\\
		\qquad \qquad \qquad \qquad \qquad \qquad \; \; \text{if } \left( \beta_{kj}=1 \right) \text{ or } \left( \beta_{kj}<1 \text{ and } 2 r_0  \frac{ cos\theta-\beta_{kj} }{ 1-\beta_{kj}^2 } < z < 2 r_0  \frac{ cos\theta+\beta_{kj} }{ 1-\beta_{kj}^2 } \right) 
		\\
		0, 
		\qquad \qquad \qquad \qquad \qquad \qquad \qquad \qquad \qquad \quad \quad \quad \quad \; \; \; \; 
		\text{if } \left( \beta_{kj}<1 \text{ and } z \le 2 r_0  \frac{ cos\theta-\beta_{kj} }{ 1-\beta_{kj}^2 } \right) 
		\\
		\pi \frac{2(z-r_0\cos\theta)}{\beta_{kj}^2}, 
		\qquad \qquad \qquad \qquad \qquad \qquad \qquad \qquad \qquad \quad \!
		\text{if } \left( \beta_{kj}<1 \text{ and } 2 r_0  \frac{ cos\theta+\beta_{kj} }{ 1-\beta_{kj}^2 } \le z \right)
	\end{cases}.\nonumber \\ 
	\label{eq:dA}
\end{IEEEeqnarray}

$\mathbb{E}[L \mid \text{tier}=k]$ is also provided in the following theorem.
\begin{Theorem} \label{Thm2}
	The mean length of the chords lying in tier-$k$ cells is as
	\begin{IEEEeqnarray}{rCl}
		\mathbb{E}[L \mid \emph{tier}=k ] = \pi \frac{ \left( \sum_{j\in\mathcal{K}} \lambda_j \beta_{jk}^2 \right)^{1/2} }{ \sum_{j\in\mathcal{K}} \lambda_j \mathcal{I}(\beta_{kj}) }, \nonumber 
	\end{IEEEeqnarray}
	where 
	\begin{IEEEeqnarray}{rCl}
		\mathcal{I}(\beta) = 
		\begin{cases}
			\frac{1}{\beta^2} \int_0^{\pi} \frac{\beta^2+1-2\cos^2\theta}{\sqrt{\beta^2-\cos^2\theta}} {\rm d}\theta,& \text{if } \beta \ge 1 \\
			\frac{1}{\beta^2} \int_{ \arccos(\beta) }^{ \pi-\arccos(\beta) } \frac{\beta^2+1-2\cos^2\theta}{\sqrt{\beta^2-\cos^2\theta}} {\rm d}\theta ,& \text{if } \beta < 1
		\end{cases}.
		\label{eq:I}	
	\end{IEEEeqnarray}
\end{Theorem}
\begin{IEEEproof}
	See \textbf{Appendix~C}.
\end{IEEEproof}

Using these results, the mean and the CCDF of the sojourn time are
\begin{IEEEeqnarray}{rCl}
	\mathbb{E}[S \mid \text{tier}=k ] &=& \frac{\pi}{v} \frac{ \left( \sum_{j\in\mathcal{K}} \lambda_j \beta_{jk}^2 \right)^{1/2} }{ \sum_{j\in\mathcal{K}} \lambda_j \mathcal{I}(\beta_{kj}) }, \label{eq:mean_sojourn_time} 
	\\
	\bar{F}_S(T \mid \text{tier}=k) &=& \mathbb{E}[L \mid \text{tier}=k] \frac{{\rm d}}{{\rm d}z} H_{\ell}( z\mid \text{tier}=k ) \Big|_{z=vT}. \label{eq:distribution_sojourn_time}
\end{IEEEeqnarray}

\subsection{Handoff Rate}
\begin{figure}
	\centering
	\includegraphics[width=.5\textwidth]{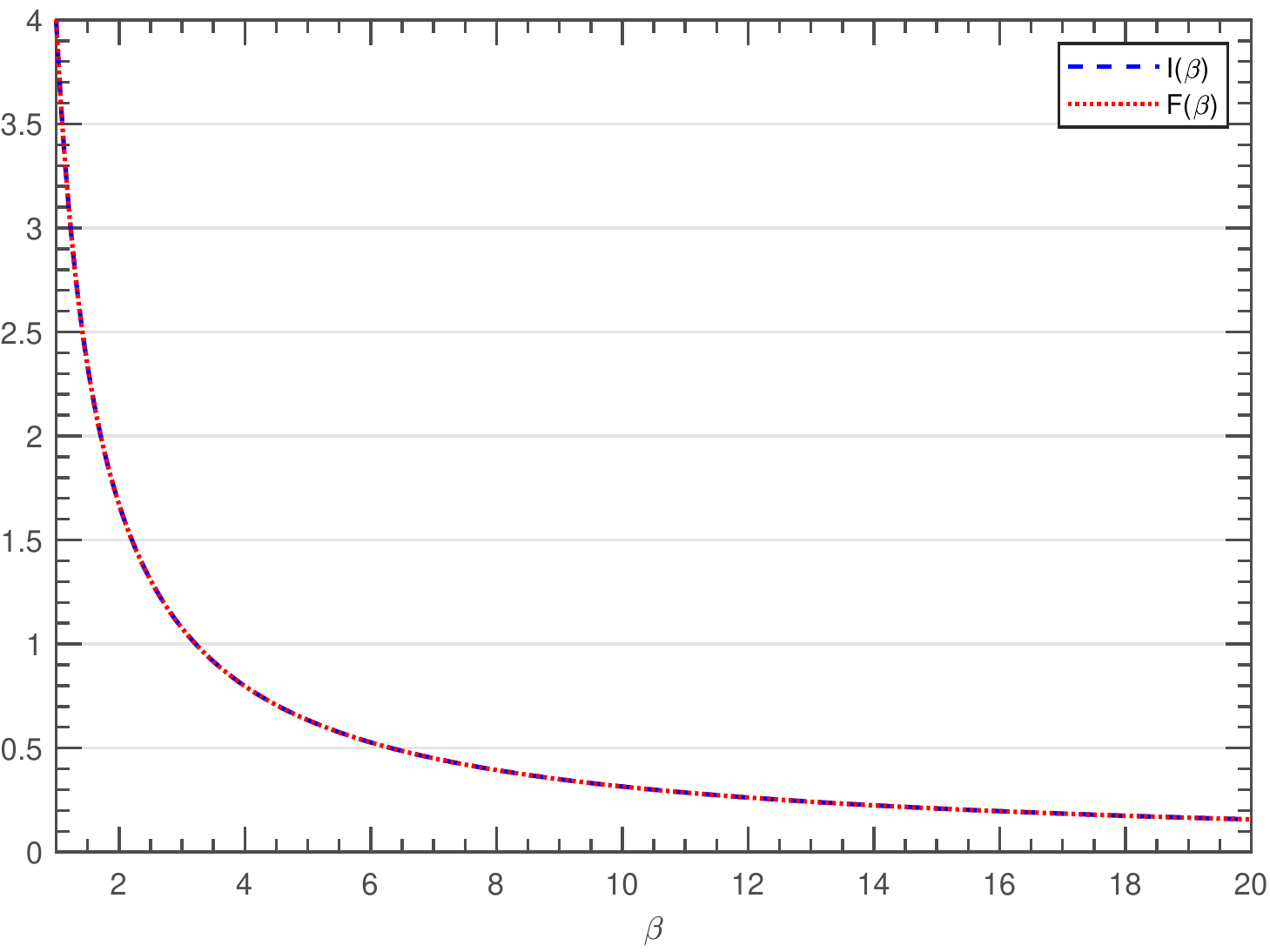}
	\caption{$\mathcal{I}(\beta)$ and $\mathcal{F}(\beta)$ for $\beta \ge 1$.}
	\label{fig:I_and_F}	
\end{figure}
In \cite{bao2015stochastic}, rates of different handoff types in multi-tier cellular networks are provided. For $k,j\in\mathcal{K}$, the type $k$-$j$ handoff rate $H_{kj}$, defined as the mean number of handoffs made from a tier-$k$ cell to a tier-$j$ cell in unit time, is\footnote{ $H_{kj}$ in \eqref{eq:kj-handoff-rate} is obtained by further simplifying the result in \cite{bao2015stochastic}. Specifically, we have used $\beta_{ij}=\frac{ \beta_{ik} }{ \beta_{jk} }$ besides $\mathcal{F}(\frac{1}{\beta})=\beta^3\mathcal{F}(\beta)$.}  
\begin{IEEEeqnarray}{rCl}
	H_{kj} = \frac{v}{\pi} \frac{ \lambda_{k}\lambda_{j}\mathcal{F}(\beta_{kj}) }{ \left( \sum_{i\in\mathcal{K}} \lambda_i \beta_{ik}^2 \right)^{3/2} },
	\label{eq:kj-handoff-rate}
\end{IEEEeqnarray}
where 
\begin{IEEEeqnarray}{rCl}
	\mathcal{F}(\beta) = \frac{1}{\beta^2} \int_0^{\pi} \sqrt{\beta^2+1-2\beta\cos\theta} {\rm d}\theta.
	\label{eq:F}
\end{IEEEeqnarray}
Therefore, 
\begin{IEEEeqnarray}{rCl}
	H_k = \frac{v}{\pi} \lambda_{k} \frac{ \sum_{j\in\mathcal{K}} \lambda_{j}\mathcal{F}(\beta_{kj}) }
	                                     { \left( \sum_{i\in\mathcal{K}} \lambda_i \beta_{ik}^2 \right)^{3/2} }
	\label{eq:H_k}
\end{IEEEeqnarray}
is the mean number of handoffs from (to) a tier-$k$ cell to (from) any other cell in the network.


In \figref{fig:I_and_F}, we can see $\mathcal{I}(\beta)=\mathcal{F}(\beta)$ for $\beta\ge 1$. Since $\mathcal{F}(\frac{1}{\beta})=\beta^3\mathcal{F}(\beta)$ and $\mathcal{I}(\frac{1}{\beta})=\beta^3\mathcal{I}(\beta)$, we can conclude $\mathcal{I}(\beta)=\mathcal{F}(\beta)$ for any $\beta>0$. Using this result, the relation between mean sojourn time and handoff rate for tier-$k$ is as follows:
\begin{IEEEeqnarray}{rCl}
	\mathbb{E}[S \mid \text{tier}=k ] = \frac{ \mathbb{P}(\text{tier}=k) }{H_k}.
	\label{eq:S_and_H}
\end{IEEEeqnarray}

An important metric in mobility analysis is the fraction of time the mobile user stays in tier-$k$ cells during a movement period since it considers both handoff rate and sojourn time. 

\begin{Corollary}
	In high velocity scenarios or mobility models with low direction switch rate, the fraction of time the mobile user stays in tier-$k$, during the movement period, is $\mathbb{P}(\emph{tier}=k)$. 
\end{Corollary}
\begin{IEEEproof}
	Let us denote the number of times that the mobile user enters a tier-$k$ cell during the movement period by $\mathcal{N}_k$, $k\in\mathcal{K}$. Also, $t_i^{(k)}$ denotes the $i$-th dwell time in the tier-$k$ cell, where $i=1,...,\mathcal{N}_k$ and $k\in\mathcal{K}$. The fraction of time that the mobile user stays in tier-$k$ can be obtained by
	\begin{IEEEeqnarray}{rCl}
		\frac{ \sum_{i=1}^{\mathcal{N}_k} t_i^{(k)} }{ \sum_{j\in\mathcal{K}}\sum_{i=1}^{\mathcal{N}_j} t_i^{(j)} } = 
		\frac{ \sum_{i=1}^{\mathcal{N}_k} t_i^{(k)} }{\mathcal{N}_k} \times \frac{\mathcal{N}_k}{ \sum_{j\in\mathcal{K}}\sum_{i=1}^{\mathcal{N}_j} t_i^{(j)} } 
		\stackrel{\text{(a)}}{=}
		\mathbb{E}[S \mid \text{tier}=k ] \times H_k \stackrel{\text{(b)}}{=} \mathbb{P}(\text{tier}=k),
		\nonumber
	\end{IEEEeqnarray}
	where (a) is obtained since we have assumed user crosses a large number of cell boundaries during a movement period, i.e. $vT$ is large compared to the average cell size. (b) also follows from \eqref{eq:S_and_H}.
\end{IEEEproof}

\subsection{Unconditional CCDF and Mean of Sojourn Time}

So far we have focused on $\bar{F}_S(T \mid \text{tier}=k)$ and $\mathbb{E}[S \mid \text{tier}=k ]$, i.e. the CCDF and average of sojourn time in a tier-$k$ cell. In the following corollaries, we derive the unconditional CCDF and mean.

\begin{Corollary}
	In high velocity scenarios or mobility models with low direction switch rate, the CCDF of sojourn time, during the movement period, can be obtained by
	\begin{IEEEeqnarray}{rCl}
		\bar{F}_S(T)= \sum_{k\in\mathcal{K}} \bar{F}_S(T \mid \emph{tier}=k) \frac{H_k}{H} \nonumber
	\end{IEEEeqnarray}
	where $H=\sum_{k\in\mathcal{K}} H_k$ is the mean number of handoffs in unit time.
\end{Corollary}
\begin{IEEEproof} Using the same notation as in the proof of \textbf{Corollary~1}, we can write
	\begin{IEEEeqnarray}{rCl}
		\bar{F}_S(T) =\mathbb{E}\left[ \mathbf{1} (S>T) \right] 
		&=& \frac{ \sum_{ k\in\mathcal{K} }\sum_{i=1}^{\mathcal{N}_k} \mathbf{1} (t_i^{(k)}>T) }
			   { \sum_{ j\in\mathcal{K} } \mathcal{N}_j } \nonumber 
		\\
		&=& \sum_{ k\in\mathcal{K} } \frac{ \sum_{i=1}^{\mathcal{N}_k} \mathbf{1} (t_i^{(k)}>T) }{ \mathcal{N}_k }
						      \times \frac{ \mathcal{N}_k }{ \sum_{ j\in\mathcal{K} } \sum_{i=1}^{\mathcal{N}_j} t_i^{(j)} } 
						      \times \frac{ \sum_{ j\in\mathcal{K} } \sum_{i=1}^{\mathcal{N}_j} t_i^{(j)} }{ \sum_{ j\in\mathcal{K} }\mathcal{N}_j }.\nonumber
	\end{IEEEeqnarray}
\end{IEEEproof} 

\begin{Corollary}
	The mean sojourn time during a movement period can be obtained by
	\begin{IEEEeqnarray}{rCl}
		\mathbb{E}\left[ S \right]=\frac{1}{H} =
		\frac{\pi}{v} \left( \sum_{k\in\mathcal{K}} \lambda_{k} \frac{ \sum_{j\in\mathcal{K}} \lambda_{j}\mathcal{F}(\beta_{kj}) }
		{ \left( \sum_{i\in\mathcal{K}} \lambda_i \beta_{ik}^2 \right)^{3/2} } \right)^{-1}. \nonumber
	\end{IEEEeqnarray}
\end{Corollary}
\begin{IEEEproof}
	The mean sojourn time can be obtained by following the same approach as in proof of \textbf{Corollary~2}.
\end{IEEEproof} 

\subsection{Effect of Network Parameters}
In this subsection, we study the effect user velocity, transmit power, bias factor, and BS intensity on the distribution and mean of the sojourn time. 

\begin{Proposition}
	The CCDF and the mean of the sojourn time decrease as the mobile user's velocity increases.
\end{Proposition}
\begin{IEEEproof}
	This can be understood from the definition of the sojourn time. (This can also be proven from the derived analytical results.)
\end{IEEEproof}

\begin{Proposition}
	In multi-tier networks, the sojourn time of $k$-th tier increases as transmit power or bias factor of tier-$k$ increases, while sojourn time in other tiers decreases. In single-tier networks, sojourn time is independent of transmit power and bias factor. 
\end{Proposition}
\begin{IEEEproof}
	Assume that a user at location $\text{y}\in\mathbb{R}^2$ is served by a tier-$k$ BS at $x$, i.e, $\text{y}$ is in the cell of $x$. Therefore, from \eqref{eq:association-policy}, we have
	\begin{IEEEeqnarray}{rCl}
		B_k P_k \|\text{y}-x\|^{-\alpha} &\ge& B_k P_k \|\text{y}-z\|^{-\alpha}, \qquad z\in\Phi_{k}, \label{Prop:eq1} \\
		B_k P_k \|\text{y}-x\|^{-\alpha} &\ge& B_j P_j \|\text{y}-z\|^{-\alpha}, \qquad 
		j\in\mathcal{K}\setminus\{k\}, \text{ and } z\in\cup_{j\in\mathcal{K}\setminus\{k\}}\Phi_j. \label{Prop:eq2}
	\end{IEEEeqnarray}   
	From these equations, when $B'_k P'_k \ge  B_kP_k$, we obtain 
		\begin{IEEEeqnarray}{rCl}
		B'_k P'_k \|\text{y}-x\|^{-\alpha} &\ge& B'_k P'_k \|\text{y}-z\|^{-\alpha}, \qquad z\in\Phi_{k}, \nonumber \\
		B'_k P'_k \|\text{y}-x\|^{-\alpha} &\ge& B_j P_j \|\text{y}-z\|^{-\alpha}, \qquad 
		j\in\mathcal{K}\setminus\{k\}, \text{ and } z\in\cup_{j\in\mathcal{K}\setminus\{k\}}\Phi_j,  \nonumber
	\end{IEEEeqnarray} 
	i.e. $\text{y}$ is still in the cell of $x$ after increasing $B_kP_k$. Therefore, in multi-tier networks, the size of the $k$-th tier cells increases as transmit power or bias factor of tier-$k$ increases. Similarly, we can show that the size of other tiers' cells decreases with increasing transmit power or bias factor of tier-$k$. On the other hand, in single-tier networks, according to \eqref{Prop:eq1}, cell sizes are independent of transmit power and bias factor. Finally, using these results and the fact that the sojourn time is directly proportional to the size of cells, we can obtain \textbf{Proposition~3}.
\end{IEEEproof}

\begin{Proposition}
	In multi-tier networks, when the BS intensity of tier-$k$ increases, the sojourn time for other tiers decreases. 
\end{Proposition}
\begin{IEEEproof}
	According to the superposition property of PPP \cite{haenggi2012stochastic}, increasing $\lambda_k$ to $\lambda'_k$ is similar to adding a new tier of BSs with intensity $\lambda'_k-\lambda_k$, transmission power $P_k$, and bias factor $B_k$. Therefore, the size of cells of other tiers decreases when the BS intensity of $k$-th tier increases.   
\end{IEEEproof}

\section{Numerical and Simulation Results}

\subsection{Distribution of Sojourn Time}
\begin{figure}
	\parbox[c]{.5\textwidth}{%
		\centerline{\subfigure[CCDF of $\tilde{S}$.]{
			\includegraphics[width=.5\textwidth]{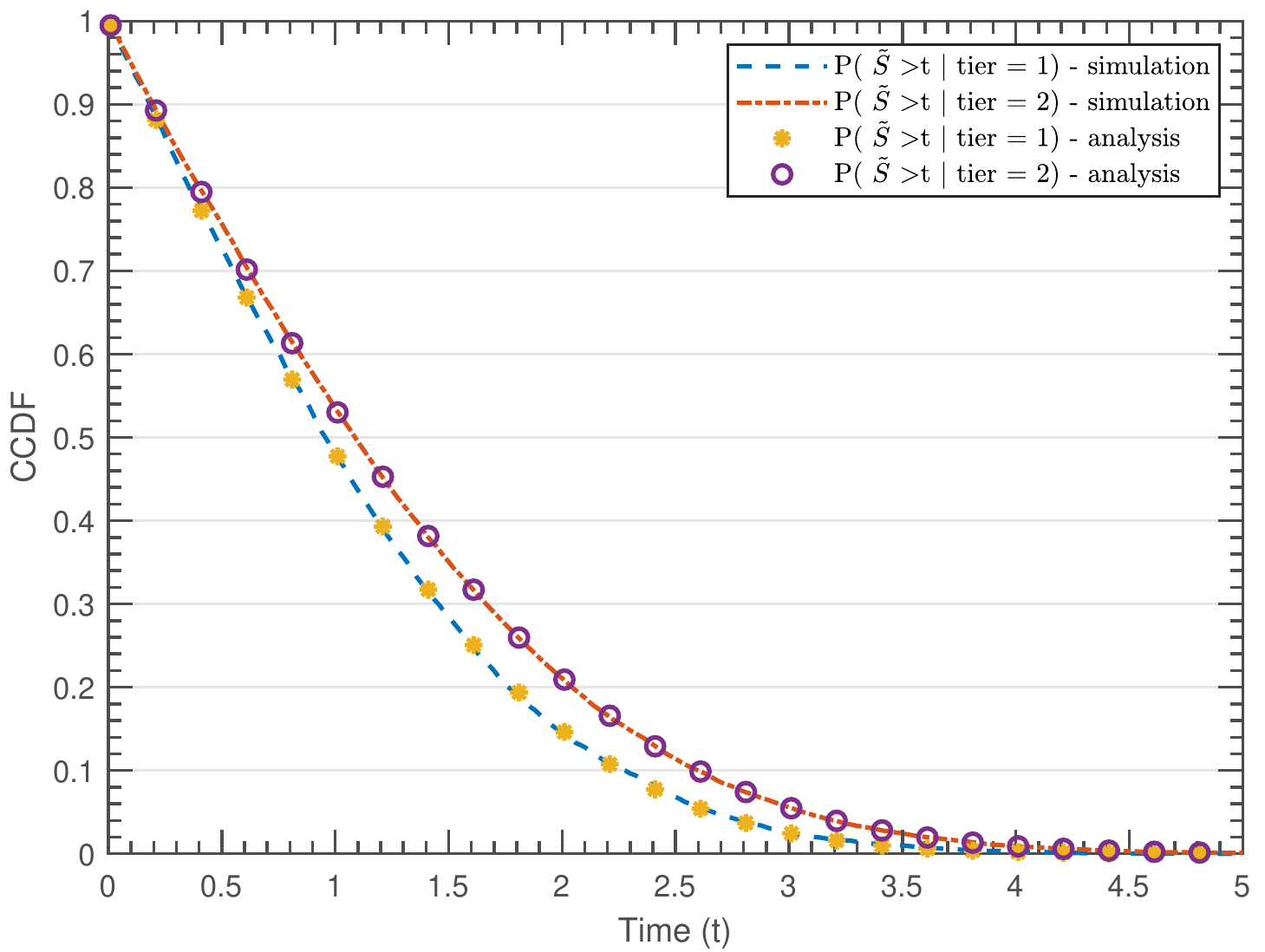}}}}
	\parbox[c]{.5\textwidth}{%
		\centerline{\subfigure[CCDF of $S$.]{
			\includegraphics[width=.5\textwidth]{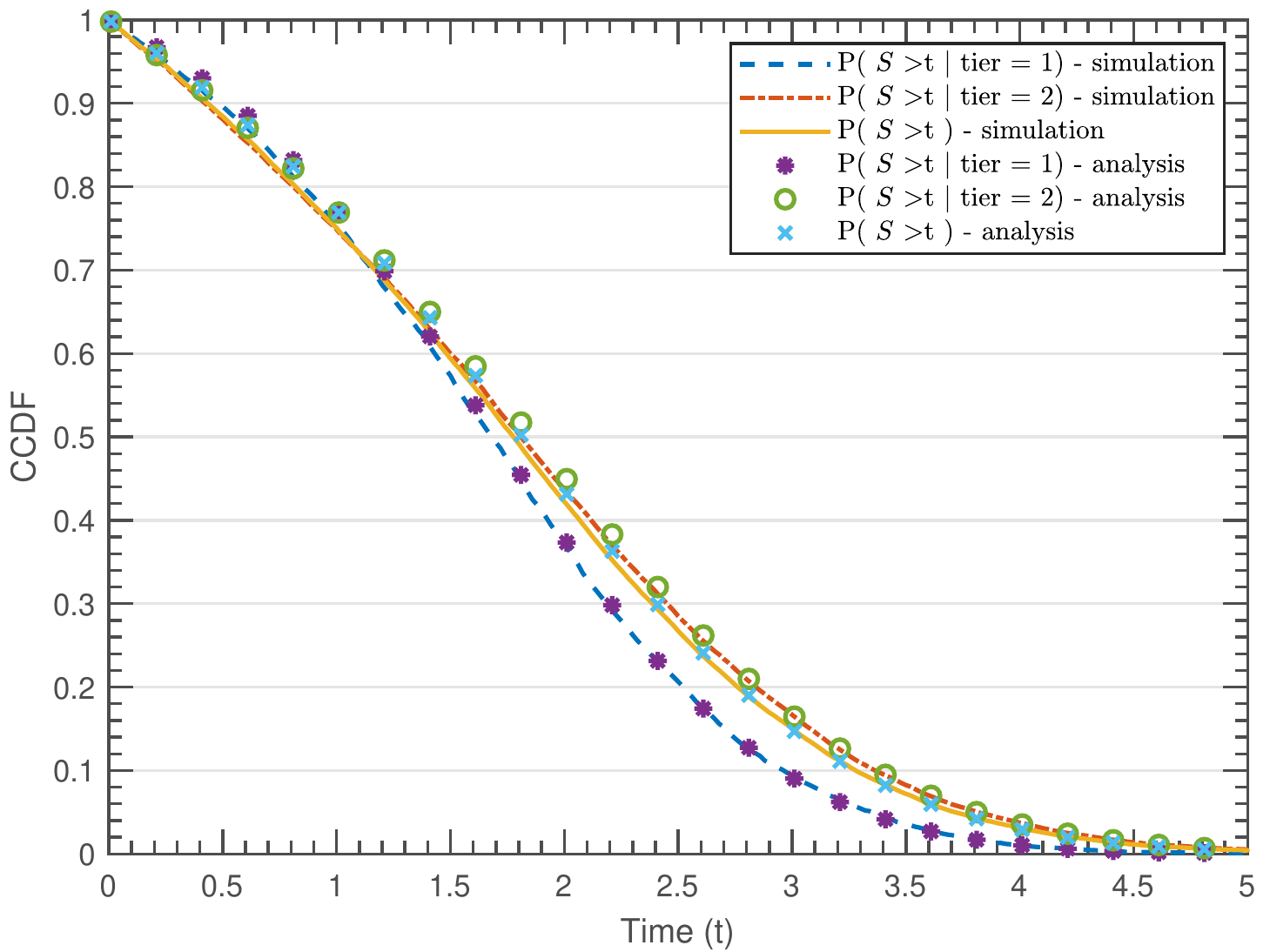}}}}
	\caption{Distribution of $\tilde{S}$ and $S$ in a two-tier cellular network (for $\lambda_1=0.002$, $\lambda_2=0.005$, $\beta_{12}=\left(\frac{1}{2}\right)^{1/4}$, and $v=5$).}
	\label{fig:CCDF}
\end{figure}
For a two-tier cellular network, in \figref{fig:CCDF}(a), the distribution of the sojourn time $\tilde{S}$ in the cell where the connection is initiated is illustrated for tier-$k$, $k\in\{1,2\}$. In \figref{fig:CCDF}(b), the distribution of (conditional and unconditional) $S$, for this network, is provided. As can be seen, the simulation results match the derived analytical results. According to \figref{fig:CCDF}, at high velocities, the sojourn time for tier-$k$ stochastically dominates the sojourn time of tier-$j$ when $B_kP_k>B_jP_j$, i.e. $\beta_{kj}>1$.

In practice, when a mobile user crosses a cell boundary, it starts a Time to Trigger (TTT) timer. The mobile user does not make a handoff to the new BS, if it leaves the new BS's cell before the end of TTT timer \cite{Xu2017}. The derived results for handoff rates, in the literature, usually assume TTT is 0, i.e. the handoff rates provided (in \cite{lin2013} and \cite{bao2015stochastic} for example) are actually the mean number of intersections between the user trajectory and cell boundaries per unit time. In practice, the handoff rate for tier-$k$ is $H_k \mathbb{P}(S>\text{TTT} \mid \text{tier}=k) = H_k \bar{F}_S(\text{TTT} \mid \text{tier}=k)$, where $H_k$ is given in \eqref{eq:H_k}. When the network parameters are as in \figref{fig:CCDF}, $H_1=0.13$ and $H_2=0.41$. For this network, with $\text{TTT}=0.2$, the handoff rate for tier-one is 0.12 and for tier-two is 0.39. Although the difference between $H_k$ and the handoff rate for these parameters is negligible, it is noticeable for high velocity scenarios.

Moreover, using the distribution of sojourn time, we can study the ping-pong rate (unnecessary handoff rate).  If, after a handoff, the time duration that the mobile user is inside the new cell be less than a threshold $T_p$, the handoff is considered unnecessary \cite{Xu2017}. Therefore, for tier-$k$, the ping-pong rate can be obtained by \cite{Xu2017}
\begin{multline}
	H_k \left( \mathbb{P}(S<T_p \mid \text{tier}=k)-\mathbb{P}(S<\text{TTT} \mid \text{tier}=k)  \right) = \\
	H_k \left( \mathbb{P}(S>\text{TTT} \mid \text{tier}=k) - \mathbb{P}(S>T_p \mid \text{tier}=k)  \right). \nonumber
\end{multline}   
For $\text{TTT}=0.2$ and $T_p = 0.5$, for tier-one, the ping-pong rate is less than $0.01$ and for tier-two, the ping-pong rate is $0.03$.    

\begin{figure}
	\centering
	\includegraphics[width=.5\textwidth]{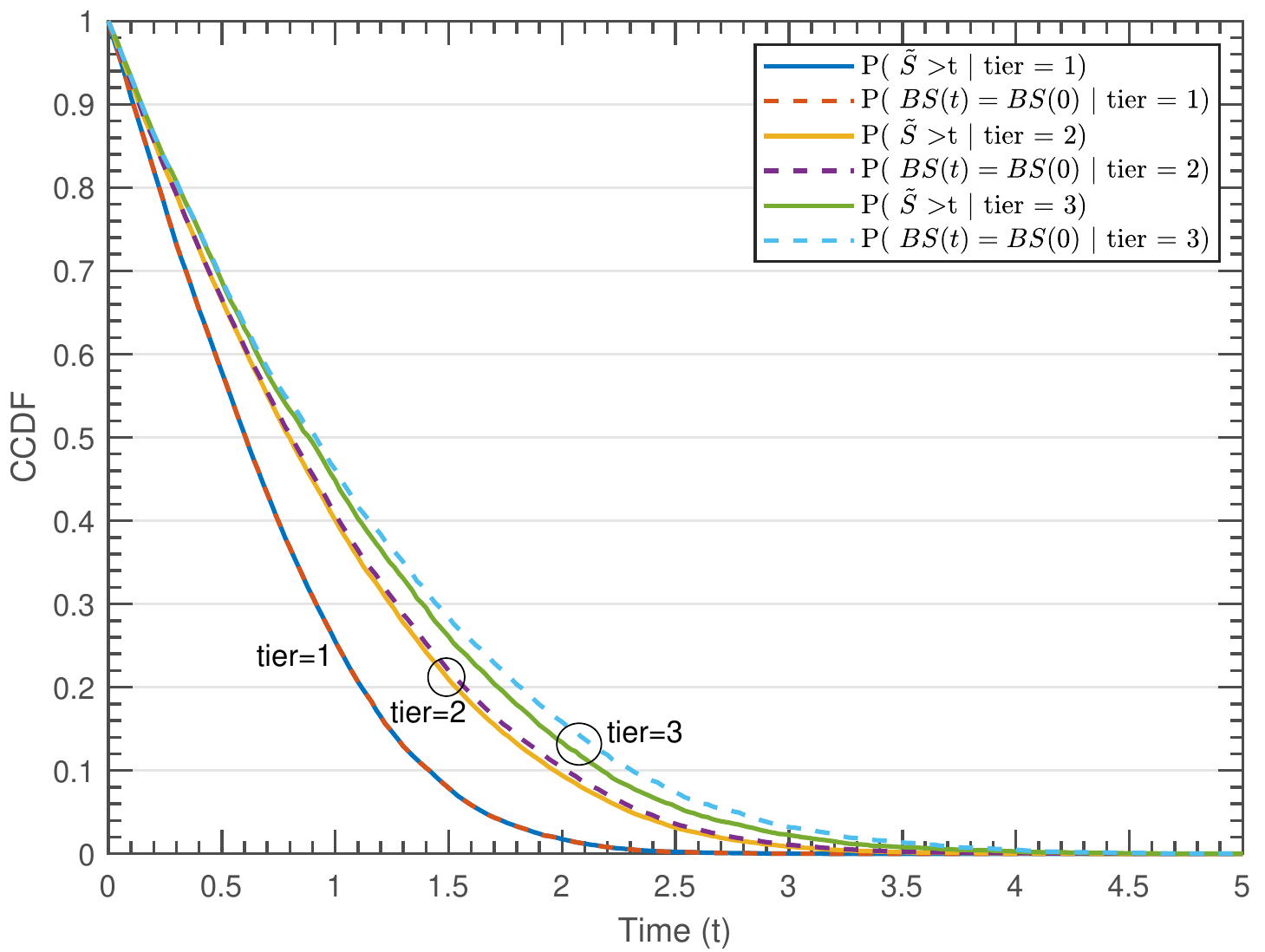}
	\caption{Sojourn time and complement of the handoff probability in a three-tier network. For tier-1, $\lambda_1=0.01$ and $B_1P_1=10$, for tier-2, $\lambda_2=0.005$ and $B_2P_2=50$, and for tier-3, $\lambda_3=0.001$ and $B_3P_3=100$. $\alpha=4$ and $v=5$. }
	\label{fig:SojournTime_and_ProbH}	
\end{figure}
As discussed earlier, we obtain the distribution of $\tilde{S}$ in multi-tier networks from \eqref{eq-sojourn-time-def}. For convex cells, \eqref{eq-sojourn-time-def} can be further simplified as \eqref{eq-sojourn-time-def-single-tier} which is the complement of the handoff probability. Therefore, we can use \eqref{eq-sojourn-time-def-single-tier} to derive the CCDF of the sojourn time in single-tier networks and also in multi-tier networks for the tier with the smallest $BP$ (multiplication of bias factor and transmission power). However, for other tiers in multi-tier networks, \eqref{eq-sojourn-time-def-single-tier} provides an upper bound for the CCDF of the sojourn time. This is also illustrated in \figref{fig:SojournTime_and_ProbH} for a three-tier network. It is worth mentioning that the gap between the CCDF of the sojourn time and its upper bound (obtained from \eqref{eq-sojourn-time-def-single-tier}) increases as the intensity of tiers with lower $BP$ increases. 

\subsection{Mean Sojourn Time}
\begin{figure}
	\centering
	\includegraphics[width=.5\textwidth]{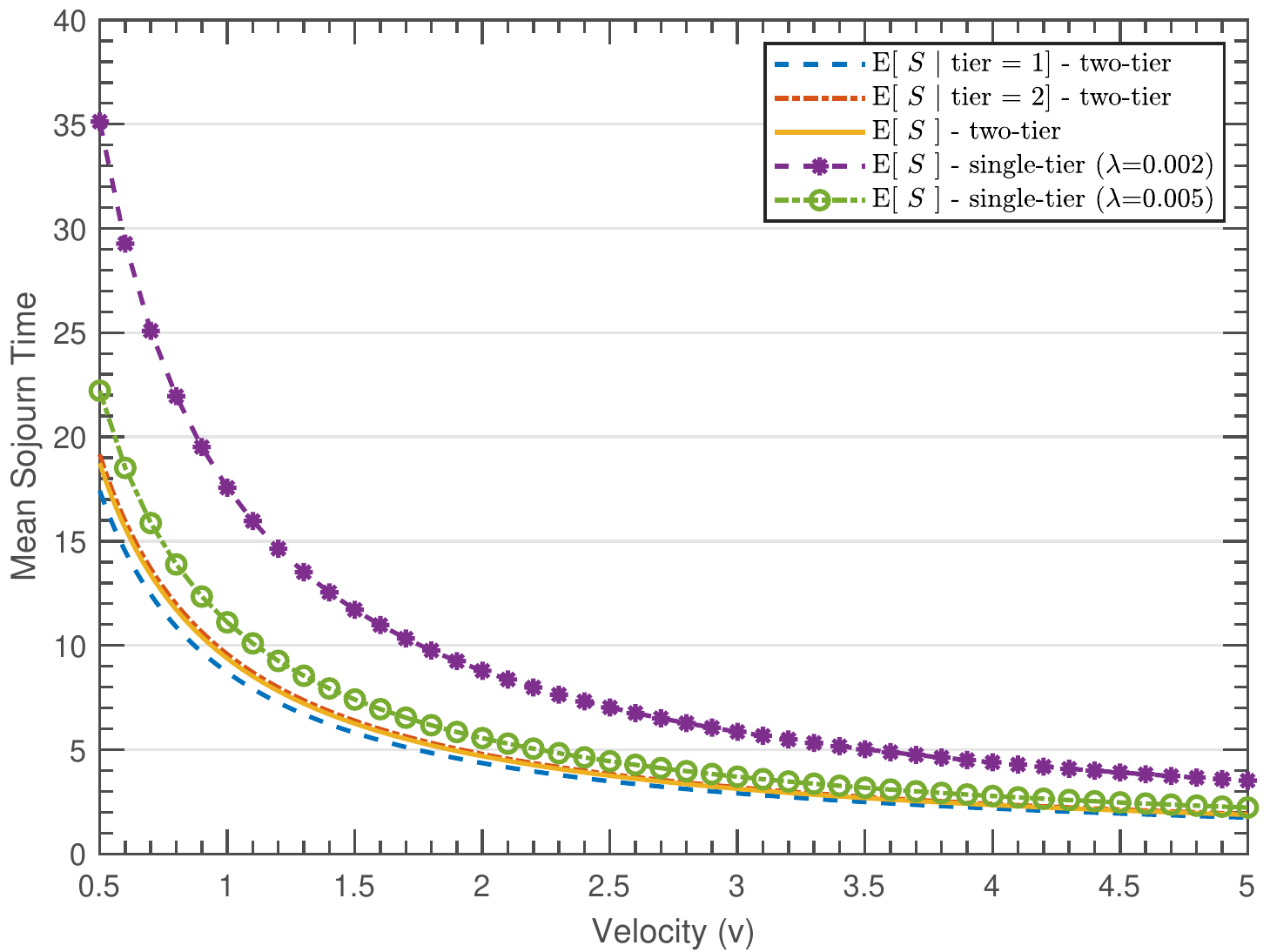}
	\caption{Mean sojourn time with respect to velocity. We have compared the results for a two-tier cellular network with $\lambda_1=0.002$, $\lambda_2=0.005$, and $\beta_{12}=\left(\frac{1}{2}\right)^{1/4}$ with two single-tier networks. }
	\label{fig:ES_v}	
\end{figure}

In \figref{fig:ES_v}, the mean sojourn time for a two-tier cellular network with $\lambda_1=0.002$ and $\lambda_2=0.005$ is illustrated as a function of velocity. We compare the results for the two-tier network with two single-tier scenarios where the mobile user is associated to only one of the tiers. When the number of tiers increases, through increased spectral reuse, users can transmit with higher data rates. However, there is more undesired overhead transmission due to the higher handoff rate (lower mean sojourn time). The sojourn time distribution is helpful in mobility management where the mobile user can skip unnecessary handoffs with a negligible spectral efficiency loss \cite{Arshad2016}.  

\begin{figure}
	\parbox[c]{.5\textwidth}{%
		\centerline{\subfigure[Effect of transmission power (or bias factor).]{
			\includegraphics[width=.5\textwidth]{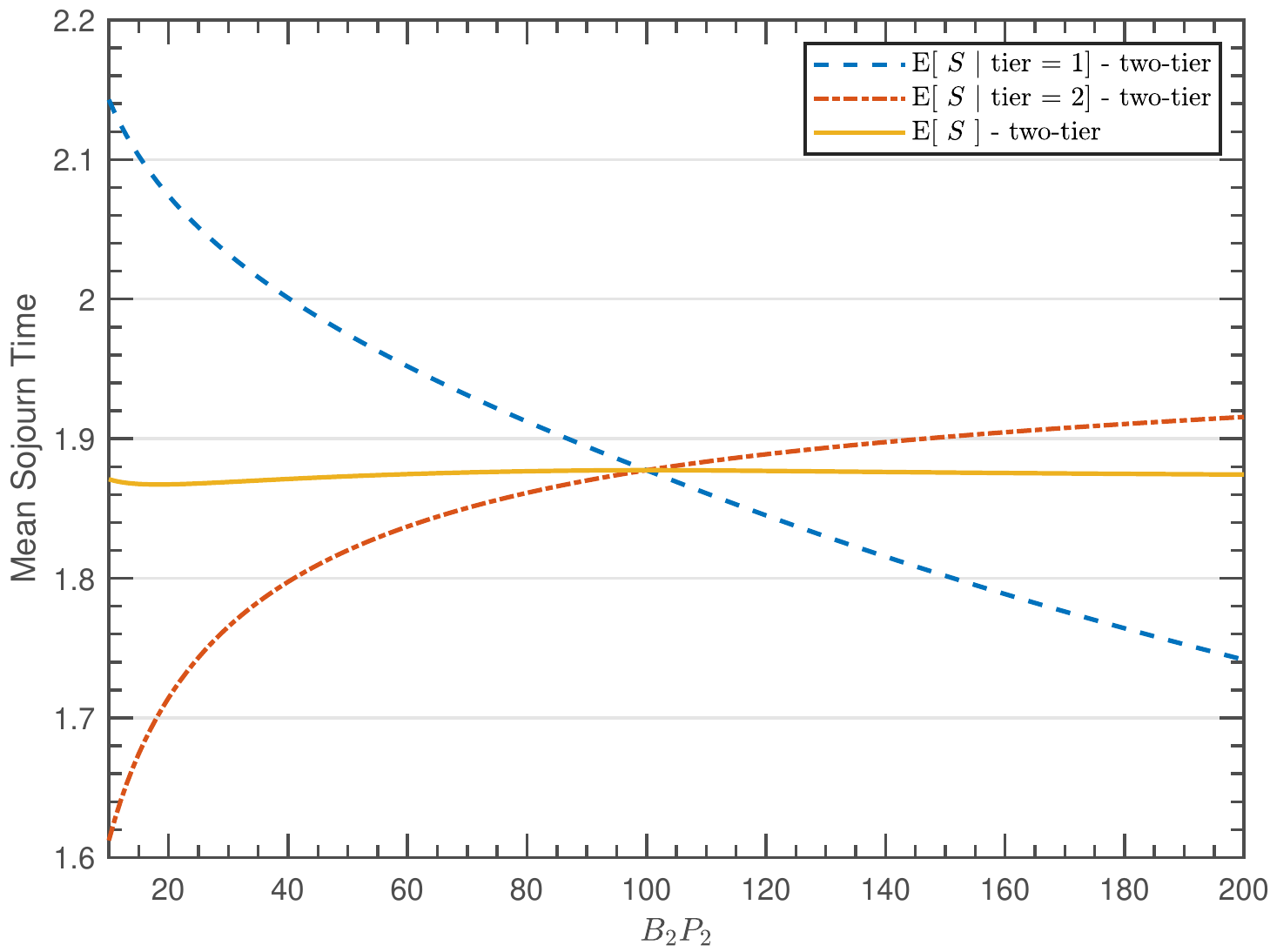}}}}
	\parbox[c]{.5\textwidth}{%
		\centerline{\subfigure[Effect of BS intensity.]{
			\includegraphics[width=.5\textwidth]{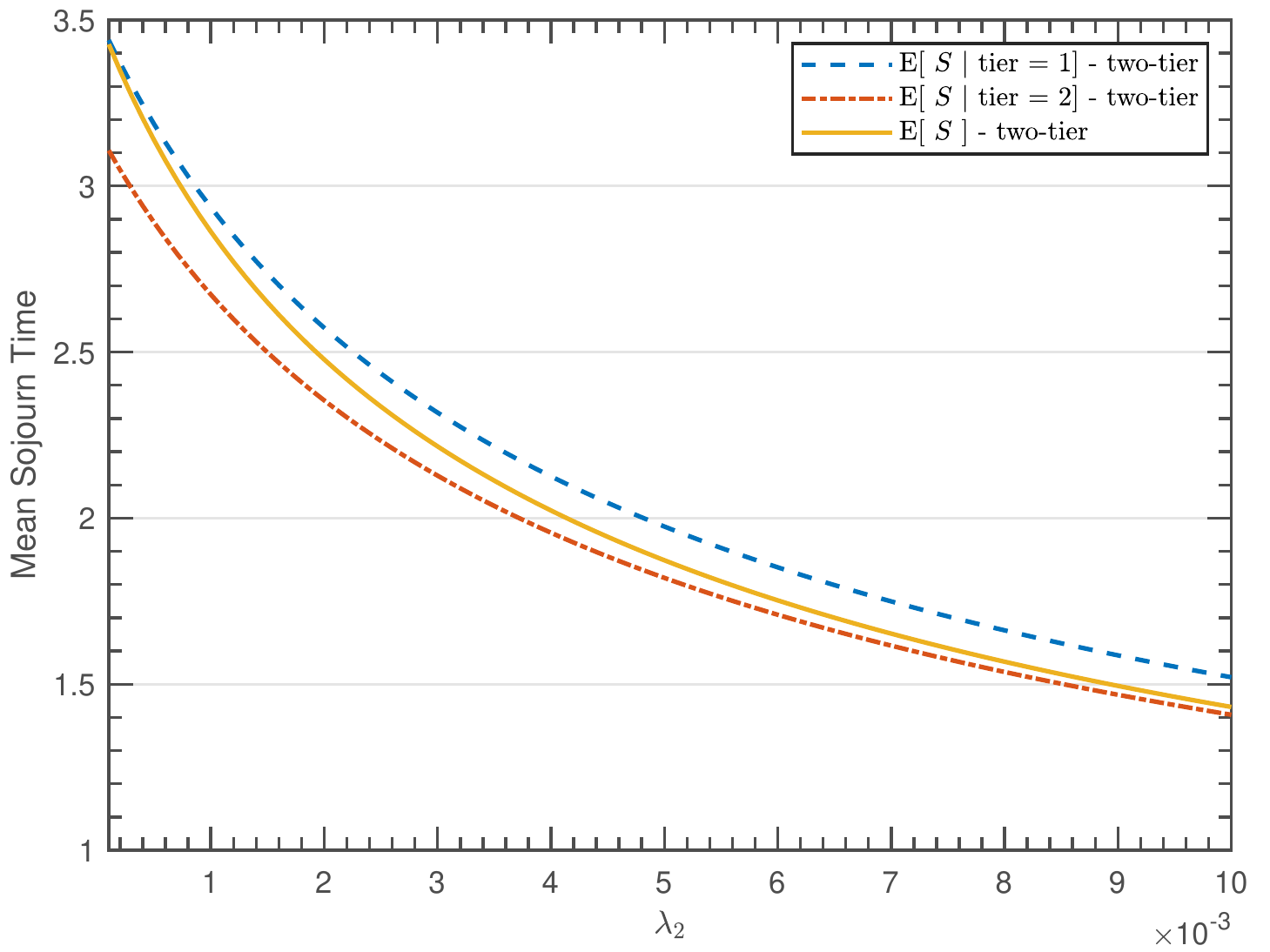}}}}
	\caption{Effect of network parameters on the mean sojourn time in a two-tier cellular network with $\lambda_1=0.002$, $B_1P_1=100$, $\alpha=4$, and $v=5$. a) Effect of increasing transmission power (or bias factor) when $\lambda_2=0.005$. b) Effect of increasing BS intensity when $B_2P_2=50$.  }
	\label{fig:ES_vs_PB}
\end{figure}
In \figref{fig:ES_vs_PB}(a), the effect of transmit power (or bias factor) on the mean sojourn time in a two-tier cellular network is illustrated. As discussed in \textbf{Proposition~3}, the mean sojourn time of tier-two increases as transmit power (or bias factor) of tier-two increases, while mean sojourn time of other tier decreases. As can be seen, the (unconditional) mean sojourn time in the network does not change with increasing transmit power or bias factor of tier-two. In \figref{fig:ES_vs_PB}(b), the effect of BS intensity on the mean sojourn time is shown. As can be seen, the mean sojourn time for all tiers decreases with increasing the BS intensity of tier-two. This is also mentioned in \textbf{Proposition~4}.

\section{Conclusion}
We have derived the distribution and mean of the sojourn time of multi-tier cellular networks. The existing works assume that a mobile user is always associated to only one of the tiers, or focus on the sojourn time in small cells (for two-tier scenario). Since in both the cases the cells are convex, the sojourn time distribution (or mean) can be easily obtained similar to single-tier scenarios by using the chord length distribution in Poisson Voronoi tessellation. However, in multi-tier networks  with maximum biased averaged received power association we need the chord length distribution in weighted Poisson Voronoi tessellation, which is not available in the literature. In this paper, we have derived the linear contact distribution function in weighted Poisson Voronoi tessellation from which we obtained the chord length distribution.

We have studied the relation between mean sojourn time and other mobility-related performance metrics. Specifically, We have shown that mean sojourn time is inversely proportional to the handoff rate. Also, the complementary cumulative distribution function of sojourn time is upper bounded by complement of the handoff probability. In addition, we have studied the impact of user velocity and network parameters on the distribution and mean of the sojourn time. The sojourn time distribution can be used to derive the ping-pong rate which is important in mobility management where the mobile user can skip unnecessary handoffs with a negligible spectral efficiency loss. Moreover, it can be used for studying channel occupancy time which can be exploited for improving resource allocation.

\section*{Appendix A: Proof of Proposition~1}
\renewcommand{\theequation}{A.\arabic{equation}}
\setcounter{equation}{0}

\begin{figure}
	\centering
	\includegraphics[width=.5\textwidth]{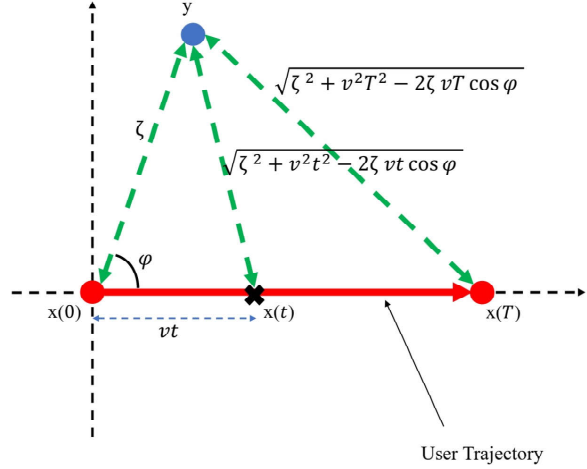}
	\caption{A geometric illustration for proof of Proposition~1.}
	\label{fig:AppendixA}	
\end{figure}

From \eqref{eq:def_green_area}, we have $ \mathcal{B}\left( \textup{x}(0),\frac{r_0}{\beta_{kj}} \right) \cup \mathcal{B}\left( \textup{x}(T),\frac{r_0(T)}{\beta_{kj}} \right) \subset \mathcal{A}_{kj}(r_0,\theta,v,T,\beta_{kj})$. To complete the proof we need to show that $\mathcal{A}_{kj}(r_0,\theta,v,T,\beta_{kj}) \subset \mathcal{B}\left( \textup{x}(0),\frac{r_0}{\beta_{kj}} \right) \cup \mathcal{B}\left( \textup{x}(T),\frac{r_0(T)}{\beta_{kj}} \right) $.

Consider a point $\text{y}\in\mathcal{B}\left( \text{x}(t),\frac{r_0(t)}{\beta_{kj}} \right)$, $0\le t\le T$. Let us represent $\text{y}$ in polar coordinates as $(\zeta,\varphi)$, where $\zeta$ is the distance between $\text{y}$ and the origin ($\text{x}(0)$) and $\varphi$ is the angle made between the line segment from the origin to $\text{y}$ and the positive $x$-axis (user's trajectory) (\figref{fig:AppendixA}). Using triangle equations, we have $\|\text{y}-\text{x}(t)\|=\sqrt{\zeta^2+v^2t^2-2\zeta vt\cos\varphi}$. Since $\text{y}\in\mathcal{B}\left( \text{x}(t),\frac{r_0(t)}{\beta_{kj}} \right)$, $0\le t\le T$,
\begin{IEEEeqnarray}{rCl}
	\zeta^2+v^2t^2-2\zeta vt\cos\varphi \le \frac{r_0(t)^2}{\beta_{kj}^2}=\frac{r_0^2+v^2t^2-2r_0 vt\cos\theta }{\beta_{kj}^2}. \nonumber
\end{IEEEeqnarray}
Rewriting the above inequality gives
\begin{IEEEeqnarray}{rCl}
	\zeta^2-\frac{r_0^2}{\beta_{kj}^2} \le \left( \frac{1}{\beta_{kj}^2}-1 \right)v^2t^2 + 2vt\left( \zeta\cos\varphi-\frac{r_0 \cos\theta }{\beta_{kj}^2} \right), \qquad 0\le t\le T. \nonumber
\end{IEEEeqnarray}
For $\beta_{kj}<1$ ($0<\frac{1}{\beta_{kj}^2}-1$), the right hand side of the above inequality is a convex function with respect to $t$. When $x\in[a,b]$, for a convex function $f$, we have $f(x)\le\max\{f(a),f(b)\}$. Using this property of convex functions yields,
\begin{IEEEeqnarray}{rCl}
	\zeta^2-\frac{r_0^2}{\beta_{kj}^2} \le \max\left\{0, \left( \frac{1}{\beta_{kj}^2}-1 \right)v^2T^2 + 2vT\left( \zeta\cos\varphi-\frac{r_0 \cos\theta }{\beta_{kj}^2} \right) \right\}. \nonumber 
\end{IEEEeqnarray}
Therefore, depending on the parameters, we have 
\begin{IEEEeqnarray}{rCl}
	\zeta^2-\frac{r_0^2}{\beta_{kj}^2} \le 0,  \quad \text{or} \quad \zeta^2-\frac{r_0^2}{\beta_{kj}^2} \le \left( \frac{1}{\beta_{kj}^2}-1 \right)v^2T^2 + 2vT\left( \zeta\cos\varphi-\frac{r_0 \cos\theta }{\beta_{kj}^2} \right).
	\label{eq:AppendixA_1}
\end{IEEEeqnarray}
We can rewrite the inequalities in \eqref{eq:AppendixA_1} as 
\begin{IEEEeqnarray}{rCl}
	\zeta^2 \le \frac{r_0^2}{\beta_{kj}^2} ,  \quad \text{or} \quad \zeta^2+v^2T^2-2\zeta vT\cos\varphi \le \frac{r_0^2+v^2T^2-2r_0vT \cos\theta}{\beta_{kj}^2}.
	\label{eq:AppendixA_2}
\end{IEEEeqnarray}
$\zeta^2 \le \frac{r_0^2}{\beta_{kj}^2}$ is equivalent to $\|\text{y}-\text{x}(0)\|\le\frac{r_0}{\beta_{kj}}$, and $\zeta^2+v^2T^2-2\zeta vT\cos\varphi \le \frac{r_0^2+v^2T^2-2r_0vT \cos\theta}{\beta_{kj}^2}$ is equivalent to $\|\text{y}-\text{x}(T)\|\le\frac{r_0(T)}{\beta_{kj}}$. Thus, $\text{y}\in \mathcal{B}\left( \textup{x}(0),\frac{r_0}{\beta_{kj}} \right) \cup \mathcal{B}\left( \textup{x}(T),\frac{r_0(T)}{\beta_{kj}} \right)$.

\section*{Appendix B: Proof of \eqref{eq:infinitesimal}}
\renewcommand{\theequation}{B.\arabic{equation}}
\setcounter{equation}{0}

Before proving \eqref{eq:infinitesimal}, we provide the Taylor series expansion of $\arccos$ and $\arcsin$. They help us to derive the final result in \eqref{eq:infinitesimal}.

The $\arcsin$ function has a Taylor expansion:
\begin{IEEEeqnarray}{rCl}
	\arcsin(x) = \sum_{n=0}^{\infty} \frac{(2n)!}{2^{2n}(n!)^2} \frac{x^{2n+1}}{2n+1},
	\label{eq:taylor_arcsin}
\end{IEEEeqnarray}
By taking derivative with respect to $x$ from both sides of \eqref{eq:taylor_arcsin}, we get
\begin{IEEEeqnarray}{rCl}
	\frac{1}{\sqrt{1-x^2}} = \sum_{n=0}^{\infty} \frac{(2n)!}{2^{2n}(n!)^2} x^{2n}.
	\label{eq:taylor_d_arcsin}
\end{IEEEeqnarray}
Using $\arccos(x) = \frac{\pi}{2} - \arcsin(x)$, we can write
\begin{IEEEeqnarray}{rCl}
	\arccos(x) = \frac{\pi}{2} - \sum_{n=0}^{\infty} \frac{(2n)!}{2^{2n}(n!)^2} \frac{x^{2n+1}}{2n+1}.
	\label{eq:taylor_arccos}
\end{IEEEeqnarray}
When ${\rm d}x \to 0$, 
\begin{IEEEeqnarray}{rCl}
	\arccos(x+{\rm d}x) &=& \frac{\pi}{2} - \sum_{n=0}^{\infty} \frac{(2n)!}{2^{2n}(n!)^2} \frac{(x+{\rm d}x)^{2n+1}}{2n+1} \nonumber 
	\\
	&=& \frac{\pi}{2} - \sum_{n=0}^{\infty} \frac{(2n)!}{2^{2n}(n!)^2} \frac{x^{2n+1}+(2n+1)x^{2n}{\rm d}x+O({\rm d}x^2)}{2n+1} \nonumber 
	\\
	&=& \frac{\pi}{2} - \sum_{n=0}^{\infty} \frac{(2n)!}{2^{2n}(n!)^2} \frac{x^{2n+1}}{2n+1} - \sum_{n=0}^{\infty} \frac{(2n)!}{2^{2n}(n!)^2} x^{2n}{\rm d}x + O({\rm d}x^2) \nonumber 
	\\
	&\stackrel{\text{(a)}}{=}& \arccos(x) - \frac{{\rm d}x}{\sqrt{1-x^2}} + O({\rm d}x^2),
	\label{eq:taylor_arccos_infinitesimal}
\end{IEEEeqnarray}
where (a) is obtained using \eqref{eq:taylor_d_arcsin} and \eqref{eq:taylor_arccos}.

From \eqref{eq:taylor_arccos_infinitesimal}, we obtain 
\begin{IEEEeqnarray}{rCl}
	\IEEEeqnarraymulticol{3}{l}{
	\arccos\left( \frac{r_0\cos\theta-vt}{\beta_{kj}r_0(t)} + \frac{\beta_{kj}^2-1}{2\beta_{kj}} \frac{v{\rm d}t}{r_0(t)} \right) = } \nonumber \\
	&& \arccos\left( \frac{r_0\cos\theta-vt}{\beta_{kj}r_0(t)} \right) - 
	\frac{ \left(\beta_{kj}^2-1\right)v{\rm d}t }{ 2\sqrt{\beta_{kj}^2r_0(t)^2-\left(vt-r_0\cos\theta \right)^2} } + O({\rm d}t^2) \stackrel{\text{(a)}}{=} \nonumber \\
	&& \pi - \arccos\left( \frac{vt-r_0\cos\theta}{\beta_{kj}r_0(t)} \right) - \frac{ \left(\beta_{kj}^2-1\right)v{\rm d}t }{ 2\sqrt{\beta_{kj}^2r_0(t)^2-\left(vt-r_0\cos\theta \right)^2} } + O({\rm d}t^2),
	\label{eq:term1}
\end{IEEEeqnarray}
where (a) follows from $\arccos(-x)=\pi-\arccos(x)$.

Using $r_0(t+{\rm d}t)=r_0(t)\left( 1 + \frac{v{\rm d}t}{r_0(t)^2}(vt-r_0\cos\theta) + O({\rm d}t^2) \right)$, when ${\rm d}t\to0$, \eqref{eq:taylor_arccos_infinitesimal}, and Taylor expansion of $(1+x)^{-1}$ also yields
\begin{IEEEeqnarray}{rCl}
	\IEEEeqnarraymulticol{3}{l}{
	\arccos\left( \frac{vt-r_0\cos\theta}{\beta_{kj}r_0(t+{\rm d}t)} + \frac{\beta_{kj}^2+1}{2\beta_{kj}} \frac{v{\rm d}t}{r_0(t+{\rm d}t)} \right) = } \nonumber \\
	&& \arccos\left( \frac{vt-r_0\cos\theta}{\beta_{kj}r_0(t)} - \frac{(vt-r_0\cos\theta)^2}{r_0(t)^2} \frac{v{\rm d}t}{\beta_{kj}r_0(t)} + \frac{\beta_{kj}^2+1}{2\beta_{kj}} \frac{v{\rm d}t}{r_0(t)} + O({\rm d}t^2) \right) = \nonumber \\
	&& \arccos\left( \frac{vt-r_0\cos\theta}{\beta_{kj}r_0(t)} \right) - 
	\frac{v{\rm d}t}{\sqrt{\beta_{kj}^2r_0(t)^2-\left(vt-r_0\cos\theta \right)^2}} \left( \frac{\beta_{kj}^2-1}{2}+1- \frac{(vt-r_0\cos\theta)^2}{r_0(t)^2} \right) \IEEEeqnarraynumspace 
	\label{eq:term2}
\end{IEEEeqnarray}
From binomial series expansion, we also get
\begin{IEEEeqnarray}{rCl}
	\IEEEeqnarraymulticol{3}{l}{
	\sqrt{ \frac{2v{\rm d}t}{\beta_{kj}}\left( r_0(t)-\frac{vt-r_0\cos\theta}{\beta_{kj}} \right) + v^2{\rm d}t^2 \left( 1-\frac{1}{\beta_{kj}^2} \right) }  \times } \nonumber \\  
	\sqrt{ \frac{2v{\rm d}t}{\beta_{kj}}\left( r_0(t)+\frac{vt-r_0\cos\theta}{\beta_{kj}} \right) - v^2{\rm d}t^2 \left( 1-\frac{1}{\beta_{kj}^2} \right) } 
	&=& \frac{2v{\rm d}t}{\beta_{kj}^2} \sqrt{ \beta_{kj}^2 r_0(t)^2-(vt-r_0\cos\theta)^2 } + O({\rm d}t^2) \nonumber \\
	\label{eq:term3}
\end{IEEEeqnarray}
Therefore, when ${\rm d}t\to0$,
\begin{IEEEeqnarray}{rCl}
	\IEEEeqnarraymulticol{3}{l}{
		\left| \mathcal{B}\left( \textup{x}(t),\frac{r_0(t)}{\beta_{kj}} \right) \setminus \mathcal{B}\left( \textup{x}(t+{\rm d}t),\frac{r_0(t+{\rm d}t)}{\beta_{kj}} \right) \right| = \pi \frac{r_0(t)^2}{\beta_{kj}^2} } \nonumber 
	\\ 
	&&-\left[ \pi - \arccos\left( \frac{vt-r_0\cos\theta}{\beta_{kj}r_0(t)} \right) - \frac{ \left(\beta_{kj}^2-1\right)v{\rm d}t }{ 2\sqrt{\beta_{kj}^2r_0(t)^2-\left(vt-r_0\cos\theta \right)^2} } \right] \frac{r_0(t)^2}{\beta_{kj}^2} \nonumber \\
	&&-\left[ \arccos\left( \frac{vt-r_0\cos\theta}{\beta_{kj}r_0(t)} \right) - 
	\frac{v{\rm d}t}{\sqrt{\beta_{kj}^2r_0(t)^2-\left(vt-r_0\cos\theta \right)^2}} \left( \frac{\beta_{kj}^2-1}{2}+1- \frac{(vt-r_0\cos\theta)^2}{r_0(t)^2} \right) \right] \nonumber \\
	&& \times \frac{r_0(t+{\rm d}t)^2}{\beta_{kj}^2} + \frac{v{\rm d}t}{\beta_{kj}^2} \sqrt{ \beta_{kj}^2 r_0(t)^2-(vt-r_0\cos\theta)^2 } + O({\rm d}t^2).
	\label{eq:term4}
\end{IEEEeqnarray}
Finally, \eqref{eq:infinitesimal} can be obtained by substituting \eqref{eq:r_t_dt} in \eqref{eq:term4}. 
 
\section*{Appendix C: Proof of Theorem 2}
\renewcommand{\theequation}{C.\arabic{equation}}
\setcounter{equation}{0}

We can derive $\mathbb{E}[L \mid \text{tier}=k]$ using \eqref{eq:def_mean_chord_length}, i.e,
\begin{IEEEeqnarray}{rCl}
	\mathbb{E}[L \mid \text{tier}=k] = \lim_{z \to 0} \frac{z}{ H_{\ell}( z\mid \text{tier}=k ) }
	\stackrel{\text{(a)}}{=} \lim_{z \to 0} \frac{1}{ \frac{{\rm d}}{{\rm d}z} H_{\ell}( z\mid \text{tier}=k ) },
	\label{eq:C1}
\end{IEEEeqnarray}
where (a) follows from L'Hospital's rule.
 
Since $\left|\mathcal{A}_{kj}(r_0,\theta,0,1,\beta_{kj})\right|=\pi \frac{r_0^2}{\beta_{kj}^2}$, we have 
\begin{IEEEeqnarray}{rCl}
	\IEEEeqnarraymulticol{3}{l}{
	\frac{{\rm d}}{{\rm d}z} H_{\ell}( z\mid \text{tier}=k ) \Big|_{z=0} } \nonumber
	\\
	&=& \frac{1}{\mathbb{P}(\text{tier}=k)} \int_0^{\infty} \int_0^{\pi} 2\lambda_k r_0 \left( \sum_{j\in\mathcal{K}} \lambda_{j} \frac{{\rm d}}{{\rm d}z}  \left|\mathcal{A}_{kj}(r_0,\theta,z,1,\beta_{kj})\right| \Big|_{z=0} \right) \nonumber 
    \exp\left\{ -\sum_{j\in\mathcal{K}} \lambda_{j} \pi \beta_{jk}^2 r_0^2 \right\} {\rm d}\theta {\rm d}r_0 \nonumber
	\\
	&=& \frac{1}{\mathbb{P}(\text{tier}=k)} \int_0^{\infty} 2\lambda_k r_0 \left( \sum_{j\in\mathcal{K}} \lambda_{j} \int_0^{\pi}  \frac{{\rm d}}{{\rm d}z}  \left|\mathcal{A}_{kj}(r_0,\theta,z,1,\beta_{kj})\right| \Big|_{z=0} {\rm d}\theta \right) \exp\left\{ -\sum_{j\in\mathcal{K}} \lambda_{j} \pi \beta_{jk}^2 r_0^2 \right\} {\rm d}r_0. \nonumber
	\\
	\label{eq:C2}
\end{IEEEeqnarray}
(Note that in the above equations we have used $\beta_{jk}=\frac{1}{\beta_{kj}}$.)

By setting $z=0$ in \eqref{eq:dA}, we get
\begin{IEEEeqnarray}{rCl}
	\IEEEeqnarraymulticol{3}{l}{
	\frac{{\rm d}}{{\rm d}z} |\mathcal{A}_{kj}(r_0,\theta,z,1,\beta_{kj})| \Big|_{z=0} = } \nonumber \\
	\begin{cases}
		\frac{2r_0}{\beta_{kj}^2}\left[\sqrt{\beta_{kj}^2-\cos^2\theta}-\cos\theta\arccos\left(\frac{\cos\theta}{\beta_{kj}}\right)\right], 
		\\
		\qquad \qquad \qquad \qquad \quad \quad
		\text{if } \left( \beta_{kj} \ge 1 \right)  \text{ or } \left( \beta_{kj} < 1 \text{ and } \arccos(\beta_{kj})<\theta<\pi-\arccos(\beta_{kj}) \right)
		\\
		0,  \qquad \qquad \qquad \qquad \qquad \qquad \qquad \qquad \qquad \quad \quad  \qquad \quad \!
		\text{if }  \left( \beta_{kj} < 1 \text{ and } \theta \le \arccos(\beta_{kj}) \right)
		\\
		-\frac{2\pi r_0 \cos\theta}{\beta_{kj}^2} \qquad \qquad \qquad \qquad \qquad \qquad \qquad \qquad \qquad
		\text{if } \left( \beta_{kj} < 1 \text{ and }  \pi - \arccos(\beta_{kj}) \le \theta \right)
	\end{cases}.\nonumber
	\label{eq:dA_at_0}	
\end{IEEEeqnarray}
Therefore, 
\begin{IEEEeqnarray}{rCl}
	\int_0^{\pi}  \frac{{\rm d}}{{\rm d}z}  \left|\mathcal{A}_{kj}(r_0,\theta,z,1,\beta_{kj})\right| \Big|_{z=0} {\rm d}\theta = 2r_0 \mathcal{I}\left( \beta_{kj} \right). 
	\label{eq:C3}
\end{IEEEeqnarray}
 
Substituting \eqref{eq:C3} in \eqref{eq:C2}, we obtain
\begin{IEEEeqnarray}{rCl}
	\IEEEeqnarraymulticol{3}{l}{
		\frac{{\rm d}}{{\rm d}z} H_{\ell}( z\mid \text{tier}=k ) \Big|_{z=0} } \nonumber
	\\
	&=& \frac{  \sum_{j\in\mathcal{K}} \lambda_{j} \mathcal{I}\left( \beta_{kj} \right)  }{\mathbb{P}(\text{tier}=k)} 
	\int_0^{\infty} 4\lambda_k r_0^2 \exp\left\{ -\sum_{j\in\mathcal{K}} \lambda_{j} \pi \beta_{jk}^2 r_0^2 \right\} {\rm d}r_0 \nonumber
	\\
	&\stackrel{\text{(a)}}{=}& \frac{  \sum_{j\in\mathcal{K}} \lambda_{j} \mathcal{I}\left( \beta_{kj} \right)  }
							        { \pi \left( \sum_{j\in\mathcal{K}} \lambda_{j} \beta_{jk}^2 \right)^{1/2}},
	\label{eq:C4}
\end{IEEEeqnarray}
where (a) follows from change of variable $\sum_{j\in\mathcal{K}} \lambda_{j} \pi \beta_{jk}^2 r_0^2=t$. Finally, \textbf{Theorem~2} is derived by substituting \eqref{eq:C4} in \eqref{eq:C1}.

\IEEEpeerreviewmaketitle
\bibliographystyle{IEEEtran}
\bibliography{IEEEabrv,Bibliography}

\end{document}